\documentclass{aa}
\usepackage{graphicx}
\usepackage{tabularx}
\usepackage{rotating}
\usepackage{txfonts}
\usepackage{comment}
\usepackage{lscape}
\usepackage{subfigure}
\usepackage{caption}
\usepackage{float}
\usepackage{placeins}

\usepackage{ulem}
\usepackage{threeparttable}

\newcommand{\mjypb}{mJy~beam$^{-1}$}
\newcommand{\hii}{H\,\textsc{i}}
\newcommand{\hi}{H\,\textsc{i}\ 21 cm}
\newcommand{\apx}{$\sim$}
\newcommand{\eg}[1]{\citep[e.g.][]{#1}}
\newcommand{\kmps}{km~s$^{-1}$}
\newcommand{\p}[1]{$^{-#1}$}
\newcommand{\pp}[1]{$^{#1}$}

\newcommand{\tspin}{{T$_{\rm spin}$}}
\newcommand{\nhi}{$N_{\rm\hii}$}
\newcommand{\wph}{W~Hz\p{1}}

\begin{document}

   \title{Redshift evolution of the \hii\ detection rate\\ in radio-loud active galactic nuclei}


   \author{Suma Murthy\inst{1,2,3},
          Raffaella Morganti\inst{2,1},
          Nissim Kanekar\inst{4}, and
          Tom Oosterloo\inst{2,1}
         }

   \institute{Kapteyn Astronomical Institute, University of Groningen, Landleven 12, 9747 AD Groningen, The Netherlands \\
             \email{murthy@jive.eu}
         \and
             ASTRON, the Netherlands Institute for Radio Astronomy, Oude Hoogeveensedijk 4, 7991 PD Dwingeloo, The Netherlands.
        \and
            Joint Institute for VLBI ERIC, Oude Hoogeveensedijk 4, 7991 PD Dwingeloo, The Netherlands.
        \and 
    National Centre for Radio Astrophysics, Tata Institute of Fundamental Research, Pune University, Pune 411007, India}

   \date{Received 29 October 2021, accepted 30 December 2021}


 \abstract{
 We present a search for associated \hi\ absorption in a sample of 29 radio-loud active galactic nuclei (AGNs) at $0.7 < z < 1$, carried out with the upgraded Giant Metrewave Radio Telescope. We detect \hi\ absorption against none of our target AGNs, obtaining $3\sigma$ upper limits to the optical depth of $\lesssim 1$\% per 50~km~s$^{-1}$ channel. The radio luminosity of our sources is lower than that of most AGNs searched for \hi\ absorption at similar redshifts in the literature, and, for all targets except two, the UV luminosity is below the threshold $10^{23}$~W~Hz$^{-1}$, above which the \hii\ in the AGN environment has been suggested to be completely ionised. We stacked the \hi\ spectra to obtain a more stringent limit of $\approx 0.17$\% per 50~km~s$^{-1}$ channel on the average \hi\ optical depth of the sample. The sample is dominated by extended radio sources,  24 of which are extended on scales of tens of kiloparsecs. Including similar extended sources at $0.7 < z < 1.0$ from the literature, and comparing with a low-$z$ sample of extended radio sources, we find statistically significant ($\approx 3\sigma$) evidence that the strength of \hi\ absorption towards extended radio sources is weaker at $0.7<z<1.0$ than at $z < 0.25$, with a lower detection rate of \hi\ absorption at $0.7 < z < 1.0$. Redshift evolution in the physical conditions of \hii\ is the likely cause of the weaker associated \hi\ absorption at high redshifts, due to either a low \hii\ column density or a high spin temperature in high-$z$ AGN environments. }


\titlerunning{Redshift evolution of the \hii\ detection rate}
\authorrunning{Murthy et al.}

\maketitle

\section{Introduction} \label{sec:intro}

To understand the role of active galactic nuclei (AGNs) and, in particular, that of radio-loud AGNs in the evolution of galaxies, it is important to understand the mass of, and physical conditions in, the neutral atomic gas in AGN environments. \hi\ absorption is a good probe to study atomic hydrogen (\hii) in such systems as it can trace cold gas at high spatial resolution and can also help to constrain the kinematics of the absorbing gas. Furthermore, such \hi\ absorption studies can be used to trace \hii\ in radio galaxies out to high redshifts, even at $z \gtrsim 4$ \citep[see][for a review]{Morganti18}. 

At low redshifts, \hi\ studies have shown that cold gas is found more often in young and restarted radio sources and in mergers, and less often in older extended radio sources \citep[e.g.][]{Morganti01, Gupta06, Vermeulen03, Gereb15, Maccagni17, Dutta18, Dutta19}. \citet{Maccagni17} used by far the largest such \hi\ absorption study of low-$z$ AGNs, covering 248 AGNs at $z < 0.25$, to show that disturbed gas is found almost exclusively in young or restarted radio sources, while the \hii\ seen in extended radio sources mostly arises from quiescent rotating gas discs. Detailed studies of the interaction between the radio jets and the interstellar medium (ISM) have shown that the jets interact directly with the ISM and thereby increase the turbulence and also drive massive cold gas outflows \citep[e.g.][]{Morganti13, Morganti15, Morganti16, Mahony13, Murthy19}. This suggests that radio AGNs have a direct impact on their host galaxies and that the nature of the impact depends on the evolutionary stage of the AGN. A recent study, extending the work of \citet{Maccagni17} to higher redshifts, suggests that these trends continue to hold out to $z \sim 0.4$ \citep{Murthy21}.

To determine the impact of radio AGNs on galaxy evolution, \hi\ absorption studies have to be extended to higher redshifts, into and beyond the epoch of galaxy assembly \eg{Shapley11}. A number of such studies have been carried out over the last three decades \citep[e.g.][]{deWaard85, Uson91, Carilli92, Carilli98, Moore99, Rottgering99, Ishwar03, Vermeulen03, Pihlstrom03, Yan16, Curran11a, Curran19, Allison15, Aditya16, Aditya17, Aditya18a, Aditya18b, Aditya19, Aditya21}, with detections of \hi\ absorption obtained out to $z \approx 3.6$. The main clear result of these studies is that the detection rate of \hi\ absorption in radio-loud AGNs decreases as one moves to higher redshifts.

Two distinct reasons have been proposed for the observed drop in the \hii\ detection rate with redshift. The first is that the high-$z$ samples suffer from a Malmquist bias, due to which the observed high-$z$ AGNs have a systematically higher luminosity than the low-$z$ ones. A high rest-frame 1.4 GHz radio luminosity may excite \hii\ to the upper hyperfine level (increasing the effective spin temperature), while a high rest-frame 1216$\AA$ UV luminosity may ionise the gas (reducing the \hii\ column density). Both effects would reduce the \hi\ absorption strength, which could lead to the non-detection of \hi\ absorption \citep{Aditya16, Aditya18b, Curran08, Curran12}. Alternatively, redshift evolution of either the cold gas content or the \hii\ spin temperature in AGN environments could also account for the low \hi\ detection rate at high redshifts \citep{Aditya16, Aditya18b}.

In order to break the degeneracy between these two possible causes, one has to study AGN samples with similar luminosities at different redshifts, or samples with similar redshift distribution but different luminosities. Further, since the detection rate has been found to be different for compact and extended radio sources at low redshifts \citep{Morganti01, Gupta06, Gereb15, Maccagni17}, studies should also control for radio morphology. Finally, the sample size should be large enough so that the results are not dominated by Poisson errors.

\citet[][see also \citealp{Aditya16}]{Aditya18b} studied the \hi\ absorption properties of a uniformly selected sample of flat spectrum radio sources over a wide range of redshifts, finding a significant difference in the detection rate of \hi\ absorption in sources at $z>1$ and those at lower redshifts. However, the high-$z$ sources of their sample were also of high UV and radio luminosity, and it was hence not possible to break the above degeneracy. \citet{Aditya18a} extended such \hi\ studies to a sample of gigahertz-peaked spectrum (GPS) sources, obtaining the first detections of \hii\ absorption in GPS sources at $z>1$. That study found no evidence for a difference in the \hi\ detection rate at different redshifts, perhaps due to the small size of the high-$z$ sample. More recently, \citet{Grasha19} studied a large sample of GPS sources over a wide redshift range and also found a higher \hi\ detection rate at low redshifts. Unfortunately, their sample too is affected by the  luminosity bias discussed above. Indeed, very few sources at $z \gtrsim 1$ with low rest-frame UV luminosities have so far been searched for \hi\ absorption \citep[e.g.][]{Curran13a,Curran13b}. The combination of small sample size and heterogeneous radio morphology in these AGNs has meant that it has not been possible to ascertain whether the high-$z$ \hi\ detection rate in different classes of radio AGNs is the same as that at low redshifts (after controlling for the UV luminosity) or if there is indeed redshift evolution of the gas properties in AGN environments. Recent detections of high-opacity \hi\ absorbers at $z \approx 1.2$ in blind \hii\ searches \citep{Chowdhury20a} further suggest that selection effects may indeed play a role in the low detection rate of \hi\ absorption at high redshifts. However, since the number of such high-$z$ searches is small, it is still not possible to rule out redshift evolution in the gas properties.

Further, at high redshifts, most \hi\ searches have predominantly been in compact radio sources, that is, compact steep spectrum (CSS) sources, GPS sources, compact symmetric objects (CSOs), and flat spectrum sources. Thus, at high redshifts, there is also a lack of studies sampling different radio morphologies that represent different stages of the evolution of radio AGNs, which has been done at $z < 0.25$ by \citet{Gereb15} and \citet{Maccagni17}.

We present here a search for \hi\ absorption in a sample of radio AGNs at $0.7 < z < 1.0$ carried out with the upgraded Giant Metrewave Radio Telescope (GMRT).\ Our aim is to understand the rate of incidence of \hi\ absorption at these redshifts
and the redshift evolution of the \hii\ content and temperature in AGN environments.

We present the sample selection, observations, data analysis, and results in Sect.~\ref{sec:observations}, a discussion in Sect.~\ref{sec:discussion}, and  a summary in  Sect.~\ref{sec:summary}. Throughout the paper, we assume a flat $\Lambda$ cold dark matter cosmology, with H$_0$ = 67.3 km s$^{-1}$Mpc$^{-1}$, $\Omega_{\Lambda}$ = 0.685, and $\Omega_M$ = 0.315 \citep{Planck14}. For the redshift range $0.7 < z < 1.0$ covered by our sample, an angular size of $1''$ corresponds to a spatial size of $\approx 7.4-8.2$~kpc.

\section{Sample, observations, data analysis, and results}
\label{data}
\label{sec:observations}

\subsection{Sample selection}
\label{sec:sample}

The AGNs of our sample were selected by cross-matching the Faint Images of the Radio Sky at Twenty-Centimeters (FIRST; \citealt{Becker95}) survey and the Sloan Digital Sky Survey Data Release~9 (SDSS~DR9; \citealt{Ahn12}) catalogues. Initially, we observed 20 sources that had a FIRST peak flux between 50~\mjypb\ and 270~\mjypb. This was done to keep the rest-frame 1.4~GHz radio luminosity below 10\pp{27}~\wph, that is, covering lower rest-frame 1.4~GHz radio luminosities than those targeted by earlier surveys for high-$z$ \hi\ absorption \citep[e.g.][]{Aditya18a}. 

\begin{figure}
    \includegraphics[width=\linewidth]{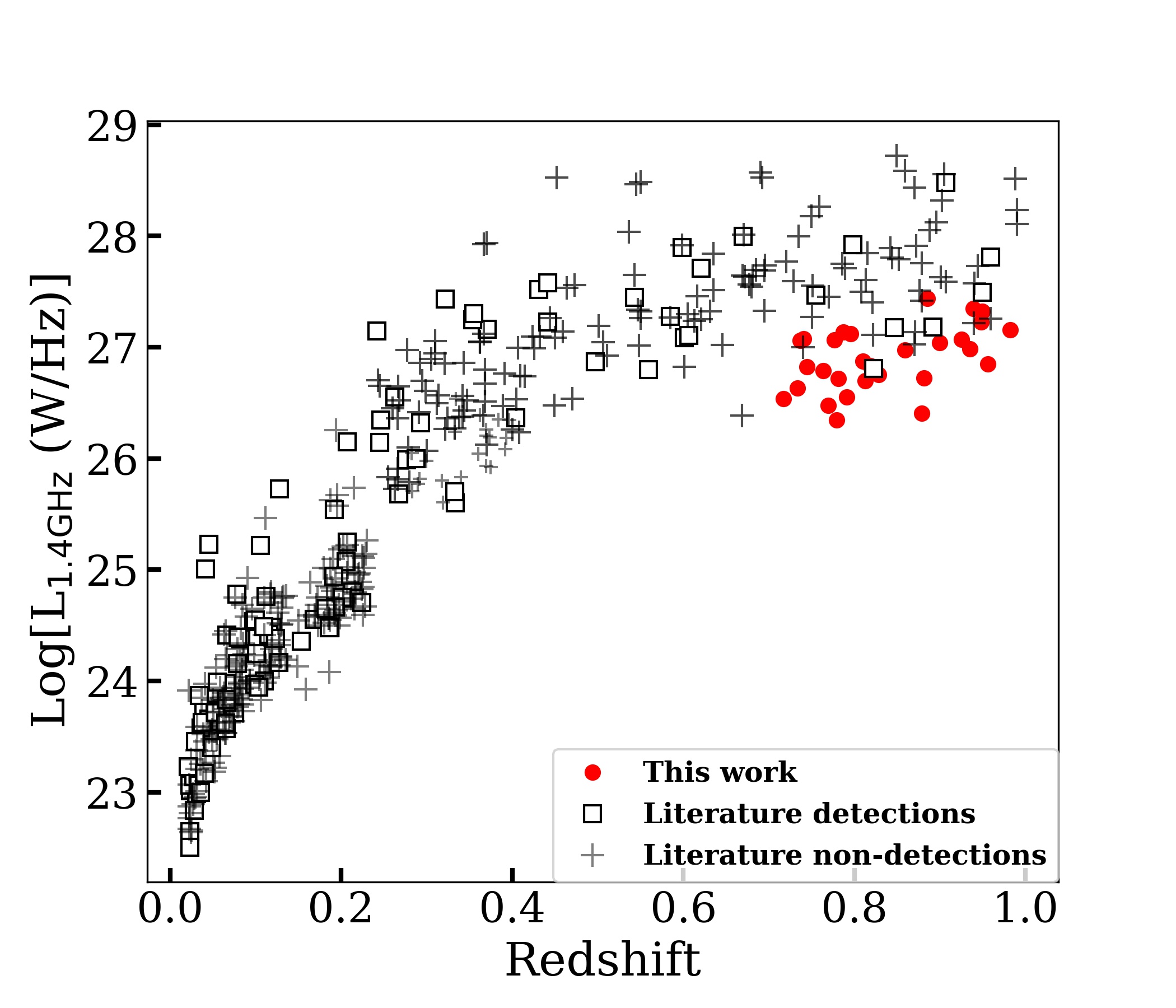}
    \caption{Distribution of radio luminosities as a function of redshift. Our sample is shown in red, and the searches reported in the literature are shown in black. At $z < 0.25$, the plot shows the sample of \citet{Maccagni17}. At $z>0.25$ the searches are from \citet{Carilli92}, \citet{Carilli98}, \citet{Pihlstrom03}, \citet{Vermeulen03}, \citet{Gupta06},  \citet{Orienti06}, \citet{Curran06,Curran11a}, \citet{Allison15}, \citet{Yan16}, \citet{Curran17}, \citet{Ostorero17}, \citet{Aditya18a,Aditya18b}, \citet{Aditya19}, \citet{Curran19}, \citet{Grasha19}, \citet{Mhaskey20}, and \citet{Murthy21}.}
    \label{fig:radio_lum}
\end{figure}

\begin{figure}
    \includegraphics[width=\linewidth]{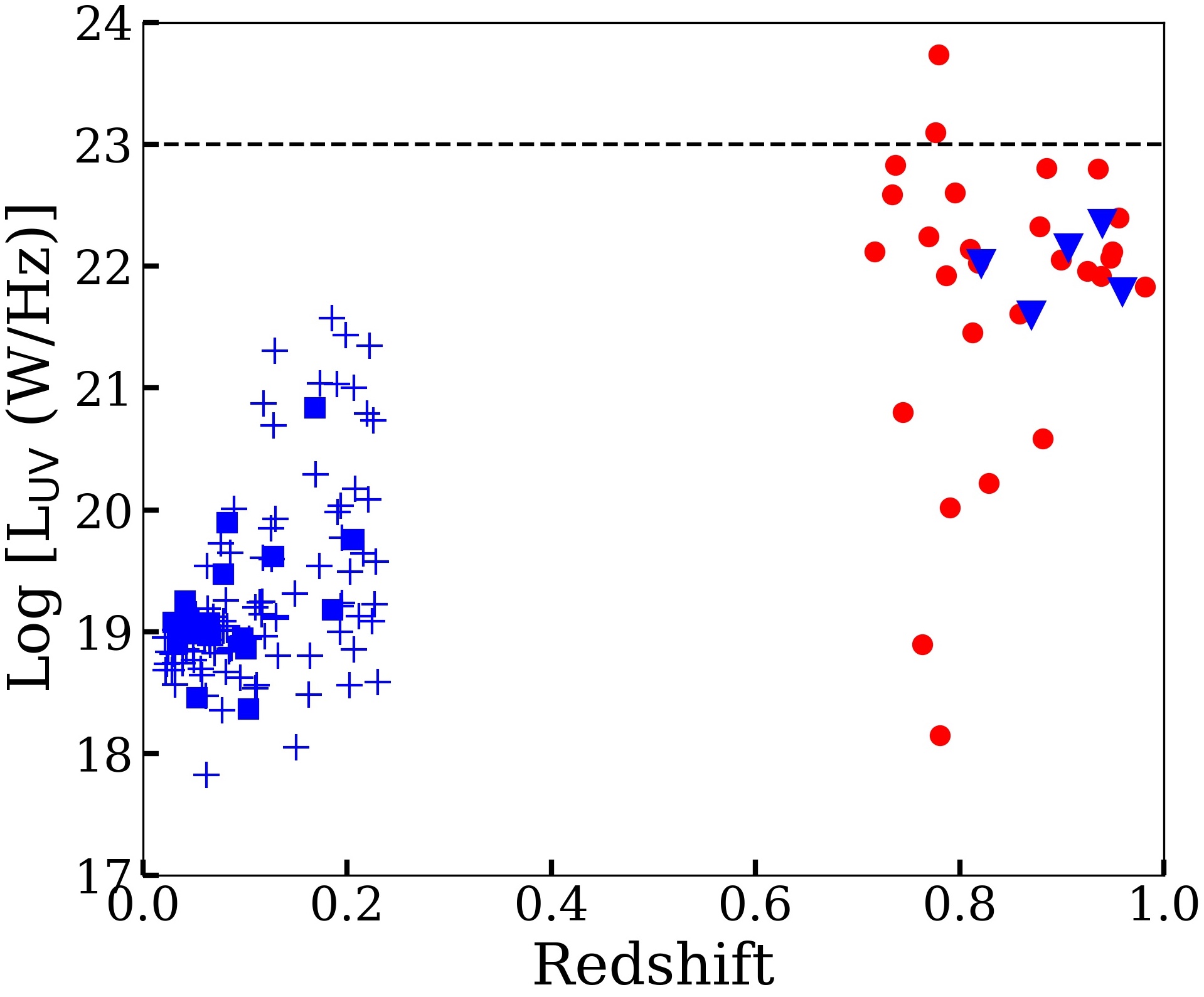}
    \caption{Comparison of the UV luminosities of the sources in our sample (red circles), the extended radio sources in the sample of \citet{Aditya19} (blue triangles), and the extended sources in \citet{Maccagni17} (blue squares: detections; blue crosses: non-detections). The dashed horizontal line marks the threshold UV luminosity $L_{UV} = 10^{23}$~\wph, above which it has been argued that the radiation from the AGN ionises the \hii\ in and around its host galaxy.}
    \label{fig:uv_lum}
\end{figure}

However, we found that most of the sources of our initial sample were classified as `quasars' by the SDSS. Thus, we observed 14 more sources, classified as `galaxies', to obtain a mix of optical AGN properties. As can be seen in Fig.~\ref{fig:radio_lum}, the typical rest-frame 1.4 GHz radio luminosity of our sample is indeed systematically lower than that of most AGNs at similar redshifts with \hi\ absorption searches. Figure~\ref{fig:uv_lum} shows that the rest-frame UV luminosity is also low, $< 10^{23}$~W~Hz$^{-1}$, for most of our targets (see Sect.~\ref{sec:luminosity} for more details). 

\subsection{Observations}

We used the Band-4 receivers of the upgraded GMRT to search for \hi\ absorption against 34 radio AGNs at $0.7<z<1.0$ (proposal IDs: 35\_079 and 36\_070; PI: Murthy). The data for three sources (SDSS\,J085448+200630, SDSS\,J021755-012150, and SDSS\,J030313-001453) were heavily affected by radio frequency interference (RFI) close to the \hi\ line frequency, and a usable spectrum could not be obtained. We hence excluded these sources from our sample. For two more sources (SDSS\,J162734+200048 and SDSS\,J100742+590809), the SDSS redshifts were found to be incorrect upon inspection of the emission lines in the optical spectra. This meant that the GMRT observations did not cover the redshifted \hi\ line frequency. After excluding the above five sources, we are left with a final sample of 29 sources. 

The on-source time for each target ranged between 10~minutes and 270~minutes, chosen to reach a 3$\sigma$ optical depth sensitivity of 1\% per 50~\kmps\ channel. A suitable flux calibrator (3C48, 3C147, or 3C286) was observed at the start of each run, and scans on the target were interleaved with those on a nearby phase calibrator. The sources of proposal 35\_079 were observed with the GMRT software backend (GSB) with a bandwidth of 16.67~MHz centred at the expected redshifted \hi\ line frequency and subdivided into 512~channels, yielding a raw spectral resolution of $\approx 12.5$~\kmps. For the remaining sources, observed in proposal 36\_070, we used the GMRT wideband backend (GWB) with a bandwidth of 25 MHz centred at the redshifted \hi\ line frequency and subdivided into 4096 channels, giving a raw spectral resolution of \apx~2.5~\kmps. The expected redshifted \hi\ line frequencies lie in the range $\approx 717 - 827$~MHz. The observational details are summarised in Table~\ref{radio_obs}\\

\subsection{Data analysis}
\label{sec:analysis}

We used the `classic' Astronomical Image Processing Software \citep[{\sc aips};][]{Greisen03} package to analyse the data. Initially, we used visual inspection to identify non-functioning antennas and data affected by RFI or other time-dependent issues and edited these data out. We then determined the antenna-based gains and bandpass shapes from the data on the flux and phase calibrators. For each source, we further improved the gain solutions through an iterative self-calibration procedure. This involved a few rounds of imaging and phase-only self-calibration. This was then followed by amplitude and phase self-calibration and imaging. We then inspected the antenna-based gains and the residual visibilities to further identify bad data and edited all such data out. We then repeated the self-calibration, imaging, and inspection procedure until the continuum image showed no further improvement and no bad data were identified in the residual visibilities. 

Next, we subtracted out the final continuum model from the calibrated visibilities, and further, for some sources, subtracted out a first- or second-order polynomial from each visibility spectrum to remove any residual continuum emission. For each source, we then made a spectral cube, shifting the $uv$ data to the systemic AGN velocity, using the optical redshift from the SDSS catalogue. In the case of the GWB data, which had a  native velocity resolution of \apx2.5~\kmps, we averaged four channels before imaging, to obtain a spectral resolution of \apx10~\kmps. 

The continuum images were made by averaging all the line channels together and have angular resolutions of \apx~$3''-6''$ and \apx~$7'' - 10''$ for {\sc robust}~$=-1$ and natural weighting, respectively. The RMS noise on the continuum images is typically a few hundred $\rm \mu$Jy beam\p{1}. The {\sc robust}$=-1$ continuum images are shown in Appendix~\ref{fig:continuum_maps}. We classified sources as resolved if the ratio of the peak flux to the integrated flux density, as measured from our highest angular-resolution image, is less than 0.9.

\subsection{\hii\ spectra}
\label{sec:spectra}

The spectral cubes were made using natural weighting, with the same restoring beam as the naturally weighted continuum images; the RMS noise on the cubes is typically $\approx 1$~\mjypb\ per \apx10~\kmps\ channel. Our final \hi\ absorption spectra were obtained by taking a cut through the AGN location in each cube. In the case of resolved radio sources, we extracted the peak continuum flux and the spectrum against the radio `core', corresponding to the location of the host galaxy as determined from the SDSS optical image. We note that this flux estimate corresponds to a region of size~$\approx 50-70$~kpc (the spatial extent of the synthesised beam at the AGN redshift), and will include not only the radio core but also extended emission within the region.
The \hi\ spectra of the 29 sources of the sample are shown in Fig.~\ref{fig:nondet_spectra}, and show no evidence for detections of \hi\ absorption.

We used the flux densities measured from the naturally-weighted continuum images to estimate the 3$\sigma$ limits on the peak \hi\ optical depth ($\tau_{3\sigma}$).  To estimate $\tau_{3\sigma}$, we used the root-mean-square (RMS) noise on the spectra smoothed to a velocity resolution of 50~\kmps\ and the core flux density. Our $3\sigma$ upper limits on the peak \hi\ optical depth lie in the range 0.3\% to 2\%, per 50~\kmps, with a median $3\sigma$ optical depth limit of 0.7\%. Finally, we used the above $\tau_{3\sigma}$ values to infer 3$\sigma$ upper limits on the \hii\ column density ($N_{\rm\hii}$), assuming a Gaussian absorption profile with a full width at half maximum (FWHM) of 50~\kmps \citep[e.g.][]{Maccagni17,Murthy21}, via the standard expression $N_{\rm \hii} = 1.82 \times 10^{18} (T_{\rm spin}/c_{\rm f}) \int (\tau \ {\rm dV})$ cm\p{2}, where $T_{\rm spin}$ is the spin temperature and $c_f$ is the covering factor. Our \hii\ column density limits further assume $T_{\rm spin} = 100$~K and $c_f = 1$. A summary of our results is provided in Table~\ref{radio_table_nondet}.

\subsection{\hii\ stacking}
\label{sec:stacking}

Stacking \hi\ absorption spectra improves the average 
\hi\ optical depth sensitivity, allowing one to probe the average cold-gas properties of the sample at lower \hii\ opacities. However, unlike the situation for \hi\ emission stacking \citep[e.g.][]{Chowdhury20b,Chowdhury21}, the associated \hi\ absorption may be offset from the optical redshifts, with the centroid either blueshifted or redshifted with respect to the systemic velocity. Stacking based on the optical redshifts may then not cause the individual \hi\ signals to align with each other; if so, the signal-to-noise ratio on the stacked spectrum may not improve $\propto 1/\sqrt{N}$, as expected for $N$ stacked spectra. However, \citet{Gereb14} stacked the spectra of all the detections of associated \hi\ absorption in their sample and found that the stacked \hi\ absorption profile was indeed centred at the systemic velocity, with a FWHM of $\approx 110$~\kmps. This suggests that, on average, the sensitivity to the average \hi\ absorption indeed improves on stacking the \hi\ signals after alignment using the optical redshifts.

\begin{figure}
    \includegraphics[width=\linewidth]{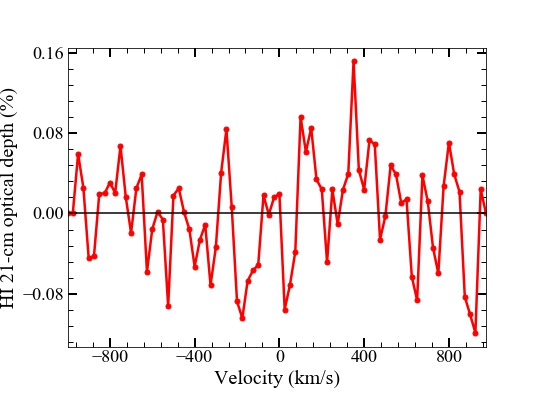}
    \caption{Stacked spectrum obtained after stacking the spectra of the 29  non-detections of \hii\ absorption. The non-detection of \hi\ absorption in the stacked spectrum yields a $3\sigma$ upper limit of $\approx 0.17$\% per 50~\kmps\ on the \hi\ opacity.}
    \label{fig:stack}
\end{figure}

We stacked the \hi\ spectra of the sources in our sample following the method of \citet[][see also \citealp{Gereb14,Maccagni17}]{Gereb13}. After aligning the \hi\ spectra in their rest frames, we converted them to optical depth units. Then we smoothed the spectra to a velocity resolution of $\approx 50$~\kmps\ and finally co-added them, using the inverse variance of each spectrum as the weight in the average. The final stacked \hi\ spectrum is shown in Fig.~\ref{fig:stack}, at a resolution of 50~\kmps, and yields no evidence of \hi\ absorption. The $3\sigma$ upper limit on the stacked \hi\ opacity is $\tau_{3\sigma} \sim 0.17$\%, per 50~\kmps\ channel.

\section{Discussion}
\label{sec:discussion}

\subsection{Radio morphology}
\label{sec:morph}

Our GMRT radio continuum images of the AGNs of the sample are shown in Fig.~\ref{fig:continuum_maps}. We find that 21 sources are resolved in the GMRT images, while eight are unresolved, at an angular resolution of $\approx 4''-6''$ (i.e. a spatial resolution of $\approx 35 - 40$~kpc at the AGN redshift). As mentioned in Sect.~ \ref{sec:intro}, earlier studies have found a higher detection rate of \hii\ absorption in compact (CSS or GPS) radio sources. Given our relatively coarse angular resolution, not all the eight sources that are unresolved in our images would be intrinsically compact sources. To investigate this further, we used flux density estimates from various surveys -- the 74~MHz VLA Low-frequency Sky Survey \citep{Lane14}, the 150~MHz TIFR-GMRT Sky Survey \citep{TGSS}, the 365~MHz Texas Survey \citep{Douglas96}, and the 1.4~GHz FIRST survey -- as well as individual measurements available in the literature to estimate their radio spectral energy distributions (SEDs). We explored whether any of the sources exhibit peaked spectra and hence are likely to be GPS or CSS, and which sources are likely to be extended on galactic scales, although unresolved in our images. 

Two of the unresolved sources (J084051+443959 and J101557+010913) are clearly GPS sources, with the SEDs showing a peak above 1~GHz \citep{Odea21}. 
A further two sources, J093150+254034 and J160325+143816, show inverted spectra, with peaks between 74~MHz and 408~MHz, while J221029+010843 shows a flattening of the spectrum at low frequencies, $\lesssim 150$~MHz; all three are likely to be CSS sources \citep{Odea21}. The last three unresolved sources (J110426+492824, J075815+441608 and J075707+273633) show a simple power-law SED  between 74~MHz and $\approx 5$~GHz, $S_\nu \propto \nu^\alpha$ with a spectral index $\lesssim -0.7$ \citep[e.g.][]{Ishwar10, Hardcastle16, Mahony16, Williams16, Jurlin21, Kukreti21}. This typical synchrotron spectrum, with no evidence for self-absorption, indicates that these sources are likely to have an extended radio morphology, with sizes greater than tens of kiloparsecs.  We thus have 24 extended radio sources and five compact ones in the sample.

\begin{figure}
    \includegraphics[width=\linewidth]{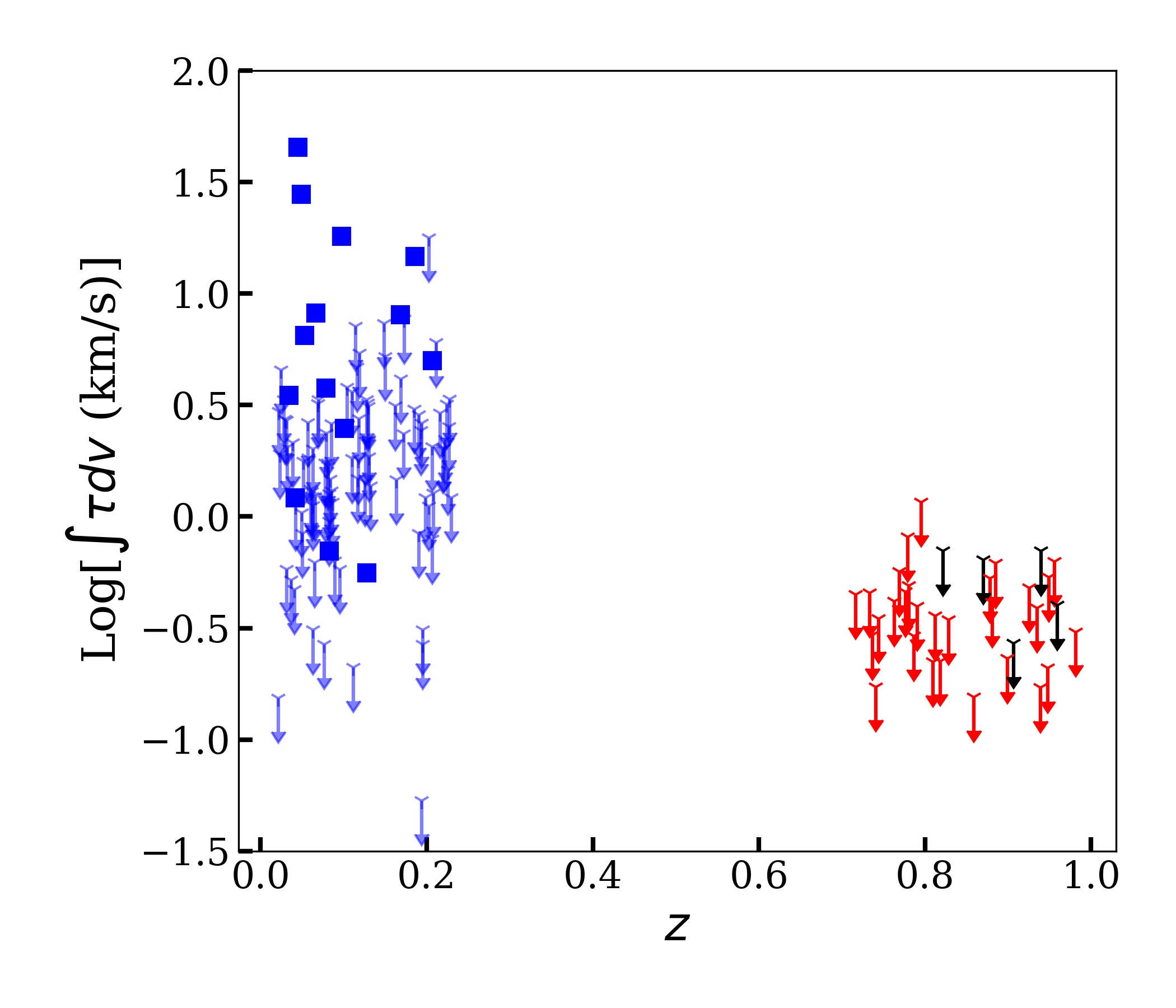}
    \caption{Comparison of the velocity-integrated \hi\ optical depth between the extended radio sources of the \citet{Maccagni17} sample (blue squares for detections and arrows for non-detections) at $z < 0.25$ and all the sources of our sample (red arrows). We have also included the extended radio sources from the study of \citet{Aditya19} (black arrows) at $0.7 < z < 1.0,$ which are combined with our sample for the analysis (see Sect. \ref{sec:redshift} for more details). The velocity-integrated \hi\ optical depth has been estimated for both samples uniformly.}
    \label{fig:int_tau}
\end{figure}

\subsection{UV and radio luminosities}
\label{sec:luminosity}

It has been suggested \citep{Curran08,Curran12} that the lack of detections of associated \hi\ absorption in high-$z$ AGNs is due to the high UV luminosity of the AGNs that have so far been searched for \hi\ absorption. \citet{Curran08} argue that there exists an AGN UV luminosity threshold of $10^{23}$~\wph, above which the \hii\ in and around the AGN host galaxy is completely ionised. 

We estimated the rest-frame 1216$\AA$ flux density ($S_{1216\AA}$) of the AGNs of our sample from  the SDSS u- and g-band magnitudes and then obtained the UV luminosity ($L_{1216\AA}$) using the expression $4 \pi D_L^2 S_{1216\AA}/(1+z)$ where $D_L$ is the AGN luminosity distance \citep[e.g.][]{Curran08}. Figure~\ref{fig:uv_lum} compares the distribution of the UV luminosities of the AGNs of our sample with that of the extended radio sources of the low-$z$ sample of \citet{Maccagni17}. The UV luminosity of our AGNs ranges from $10^{18}$~\wph\ to $10^{23.7}$~\wph\ with the median value being $1.2 \times 10^{22}$~\wph, while the median $L_{\rm UV}$ for the low-$z$ sample is $1.2 \times 10^{19}$~\wph. However, all but two of our AGNs have rest-frame UV luminosities below 10\pp{23}~\wph; our sample thus allows us to test the hypothesis of a UV luminosity threshold that results in the paucity of \hi\ absorption detections at high redshifts.

The distribution of the rest-frame 1.4 GHz radio luminosities of our sample is shown in Fig. \ref{fig:radio_lum}, along with the data for the $z<0.25$ AGNs of \citet{Maccagni17} and all the searches in the literature at $0.25 < z < 1.0$ 
\citep[e.g.][]{Aditya18a, Aditya18b, Aditya19, Yan16, Grasha19, Carilli98, Salter10}. The 1.4~GHz radio luminosity of our sources ranges from 10\pp{26.3}~\wph\ to 10\pp{27.4}~\wph. The figure shows that the typical radio luminosity of the AGNs of our sample is significantly lower than that of AGNs at $z \approx 0.5-1.0$ that have been earlier searched for \hii\ absorption: the median 1.4~GHz luminosity of our sample is \apx $6 \times 10^{26}$~\wph, nearly an order of magnitude lower than that of sources at similar redshifts in the literature (median 1.4~GHz luminosity \apx$5 \times 10^{27}$~\wph). However, the typical 1.4~GHz luminosity of our AGNs is systematically higher than that of the low-$z$ sample of \citet{Maccagni17}.

We note that, as mentioned in Sect. \ref{sec:sample}, the SDSS classification of some of the AGNs is `galaxy' while it is `quasar' for the rest, and this may suggest a possible difference in the orientation that thereby impacts the detection of \hi\ absorption. However, our sample is dominated by extended radio sources and none of these show a flat radio SED. This implies that for none of these sources the line of sight is along the radio-jet axis as a quasar 
classification would suggest. Hence we argue that orientation effects are not significantly different for the sources classified as quasars in the SDSS compared to the rest of the sample. Furthermore, \citet{Curran08} have shown that for $L_{\rm UV} < 10^{23}$~\wph, the \hi\ detection rate among the sources classified as quasars or otherwise is comparable.

\subsection{Redshift evolution of the  detection rate of \hi\ absorption}
\label{sec:redshift}

To test for evolution in the detection rate of \hi\ absorption, one has to compare with a low-redshift comparison sample with properties similar to that of the target sample. The study of \citet{Maccagni17} is the largest at $z < 0.25$ and consists of sufficient number of sources in different sub-samples (based on morphology) for a statistical analysis. Thus we compare our sample with that of \citet{Maccagni17}.

At low redshifts, \citet{Maccagni17} find a difference in the detection rates of \hi\ absorption between extended and compact radio sources: extended radio sources have a detection rate of \apx16\%, while compact radio sources have a higher detection rate of \apx30\% \citep{Gereb14, Gereb15, Maccagni17}. The lower \hi\ detection rate in extended radio sources could be due to (i)~a low covering factor or (ii)~orientation effects. The low-$z$ studies have found that most of the \hii\ arises from gas settled in a disc, in which case the chance of detecting \hii\ absorption strongly depends on the orientation of the disc with respect to the radio source. However, \citet{Gereb14} show that, even without correcting for a low covering factor, the detection rate of 16\% in extended radio AGNs is consistent with the \hi\ absorption arising from a quiescent gas disc.

Since our sample is dominated by extended radio sources, we compare the detection rate of \hi\ absorption in these sources to that in the extended radio sources of \citet{Maccagni17}. We note that it is not possible to make a similar comparison with studies at $0.25<z<0.7$ due the small number of extended radio sources searched for \hii\ absorption. 

The integrated \hi\ optical depth has been estimated uniformly for all the sources in both the samples. We note that the covering factor has not been considered for estimating the \hi\ optical depth in both high- and low-$z$ samples. The median spatial resolution for the two samples is comparable: for our sample it is \apx56 kpc while that for \citet{Maccagni17} sample is 42 kpc. Thus, we argue that a correction for covering factor is not necessary for this study.

Our \hi\ optical depth sensitivity is similar to that of \citet{Maccagni17}. Figure~\ref{fig:int_tau} compares the velocity-integrated \hii\ optical depth of all 29 sources in our sample and that of the extended radio sources in the sample of \citet{Maccagni17}. It is clear that the \hi\ optical depth sensitivity of our observations is sufficient to detect \hi\ absorbers similar to those seen by \citet{Maccagni17} at $z < 0.25$. 
Further, as mentioned in Sect.~\ref{sec:stacking}, we stacked the 29 \hi\ spectra but did not obtain a detection of the stacked \hii\ signal, with the $3\sigma$ upper limit $\tau_{3\sigma} = 0.17\%$ per 50~\kmps\ channel on the stacked \hi\ optical depth.

All of the extended radio sources of \citet{Maccagni17} have UV luminosities lower than $10^{23}$~\wph. Of the 24 extended radio sources in our sample, one has UV luminosity above the proposed cutoff limit; we exclude this source from the comparison, in order to ensure that all the sources being compared are below the UV luminosity threshold. Further, we include the extended radio sources searched for \hi\ absorption at $0.7 < z  <1.0$ reported in the literature. Most of the sources that have been searched for \hi\ absorption with sensitivity comparable to our sample in this redshift range \citep{Carilli98,Vermeulen03,Salter10,Yan16,Aditya18a, Aditya18b} are only compact radio sources. The study of \citet{Aditya19} alone consists of five extended radio sources with UV luminosity below $10^{23}$~\wph. Including these five sources, we have a final high-$z$ sample of 28 extended radio sources with no detection of \hi\ absorption at $0.7 < z < 1.0$. 

We thus obtain a detection rate of $0^{+6.6}$\% in the high-$z$ sample, where the errors are $1\sigma$ Gaussian errors taking into account small-number statistics \citep{Gehrels86}. In the low-$z$ sample, \citet{Maccagni17} obtain 17 detections of \hi\ absorption from 108 extended sources, yielding a detection rate of $15.7^{+4.8}_{-3.8}$\%. While the \hi\ detection rate appears higher in the low-$z$ sample, the relatively small size of the high-$z$ sample implies that the difference between detection rates has $< 2\sigma$ significance.

However, since the low-$z$ sample contains both detections and non-detections of \hi\ absorption, while the high-$z$ sample contains only non-detections, a proper comparison between the values of the \hi\ optical depths requires the use of survival analysis \citep[e.g.][]{Isobe86}. We hence further carried out a two-sample Peto-Prentice generalised Wilcoxon test for censored data \citep{Feigelson85, Isobe86, Isobe90} to compare the distributions of velocity-integrated \hi\ optical depths in the low-$z$ and high-$z$ samples after estimating the value uniformly for both the samples. We find that the null hypothesis that the \hi\ optical depths of the two samples are drawn from the same distribution is rejected ($p$-value $\approx 0.0029$) at $\sim 3\sigma$ significance. We thus find statistically significant evidence that the \hi\ optical depth in low-$z$ extended radio AGNs are higher than in extended AGNs at high redshifts.

\subsection{Cause of the evolution of the detection rate}

\subsubsection{UV luminosity}

All sources of the two samples have UV luminosities $< 10^{23}$~\wph. The exact effect of UV luminosity on cold gas in the host galaxy is unclear and even if not a complete ionisation of gas, the UV luminosity, may still result in a decrease in the \hi\ absorption strength. However, below $10^{23}$~\wph, such an inverse correlation between the \hi\ absorption strength and UV luminosity is not seen in earlier studies \citep[e.g.][]{Aditya18a, Aditya18b, Curran08, Curran13a}. Based on these results we argue that the effect of UV luminosity is not significant in the observed difference in the \hi\ optical depths of the two samples. A study with much larger sample size spanning a wide range of UV luminosities, possible only with deep wide-field \hii\ surveys, will be able to probe this further.

\subsubsection{Radio luminosity}

Next, we consider the contribution of rest-frame 1.4 GHz radio luminosity to the observed difference in the \hi\ optical depth in the two samples. The \hii\ spin temperature could be affected by the radiation from the AGN in the region close to the radio source \citep{Bahcall69}. Furthermore, most of the \hi\ absorption in radio AGNs arises from gas within a few kiloparsecs of the central radio source \citep{Morganti18}. Thus it is possible that high radio luminosity would have resulted in an increased \tspin\ in high-$z$ sources. In that case a higher optical depth sensitivity would be required to detect the gas of similar \nhi\ as the low-$z$ absorbers. Though the radio luminosity in our sample is lower than that of most of the sources studied so far, it is at least three orders of magnitude higher than that of the low-$z$ sources.

However, in the sample of \citet{Maccagni17}, which spans three orders of magnitude in radio luminosity, there is no dependence of the integrated \hi\ optical depth on the rest-frame 1.4 GHz radio luminosity in the entire sample. Furthermore, at $z > 0.25$, the studies reported in the literature for sources with $L_{\rm 1.4GHz} < 10^{27}$~\wph\ (the median luminosity of our sample), the integrated \hi\ optical depths do not appear to be correlated with the rest-frame 1.4 GHz luminosity \citep[see, for example,][]{Aditya18a, Aditya18b}. Thus although we cannot rule out the impact of radio luminosity on the ambient \hii, based on the empirical evidence, we argue that it is unlikely that radio luminosity could significantly affect the detection rate of \hi\ absorption.

\subsubsection{Redshift evolution of the physical conditions of HI}

It thus appears that neither the UV nor the 1.4~GHz radio luminosities of the AGNs of our sample are likely to significantly affect our low detection rates of \hi\ absorption at $z \approx 0.7-1.0$. Thus, the lower \hi\ optical depth in the high-$z$ AGN sample is likely to arise due to redshift evolution in \hii\ conditions in AGN environments at $z < 1$. This could arise either because the \hii\ spin temperature in the radio-AGN host galaxies is higher at high redshifts, as has been observed for the high-$z$ damped Lyman-$\alpha$ absorbers \citep[e.g.][]{Kanekar14}, or because the \hii\ column density is lower in high-$z$ AGN environments. An independent estimate of the \hii\ column density would be needed to disentangle the high-\tspin\ and the low-\nhi\ scenarios.

\subsection{Low-luminosity radio sources and unbiased surveys}

Low-luminosity radio sources, with rest-frame 1.4~GHz luminosities $\lesssim 10^{25}$~\wph, dominate the radio AGN population at all redshifts \citep[e.g.][]{Best05, Willott01, Simpson12, Pracy16, Slaus20}. It is hence imperative to study conditions in \hii\ in the environments of low-luminosity radio sources of different radio morphologies, and different optical and host-galaxy properties, at all redshifts to obtain a complete picture of the role of radio AGNs in the evolution of their host galaxies. Unfortunately, targeted \hi\ absorption studies of such `typical' radio AGNs at $z \gtrsim 1$ would require large amounts of observing time with the best radio telescopes today, and are hence unlikely to be feasible for large AGN samples. However, a large number of such low-luminosity AGNs would lie within a given pointing and the instantaneous \hi\ coverage of radio interferometers such as the upgraded GMRT, MeerKAT \citep{Booth09}, and the Australian Square Kilometer Array Pathfinder (ASKAP; \citealt{Johnston08}). It will hence be possible to use deep observations of individual sky fields with these telescopes to obtain unbiased \hi\ spectra of all the low-luminosity AGNs within the pointing and with redshifts such that the \hi\ line lies within the redshift coverage of the telescope. Indeed, one of the first such surveys has already yielded the detection of redshifted \hi\ absorption in a low-luminosity ($\approx 10^{25}$~\wph) AGN at $z \approx 1.2$ \citep{Chowdhury20a}. 

A further advantage of such wide-field, unbiased surveys is that they would remove the intrinsic bias against dusty sightlines that arises when targeting objects with known (optical) redshifts. The very ability to measure an optical redshift may bias one against the sightlines that are likely to produce the strongest \hi\ absorption. Such unbiased \hi\ absorption surveys have already yielded a number of new detections of redshifted \hi\ absorption \citep[e.g.][]{Allison15,Allison20,Allison21,Chowdhury20a,Mahony21}. When combined with follow-up high spatial resolution continuum imaging and studies of the optical properties of the AGNs and their host galaxies, these will provide valuable insights into the role of radio AGNs in the evolution of galaxies.

\section{Summary}
\label{sec:summary}

We report a deep GMRT search for associated \hi\ absorption in 29 radio AGNs at $0.7 < z < 1$. Our non-detections of \hi\ absorption yield $3\sigma$ optical depth limits of $\tau_{3\sigma} \lesssim 1$\% per 50~\kmps\ channel, comparable in sensitivity to low-$z$ searches for associated \hi\ absorption. We also stacked the 29 \hi\ spectra, aligning on the optical AGN redshifts, to improve the average \hi\ optical depth sensitivity; this yielded the $3\sigma$ optical depth limit $\tau_{3\sigma} < 0.17$\% on the average \hi\ optical depth of the sample. 

Most of the AGNs of our sample have UV luminosities $< 10^{23}$~\wph, which is lower than the threshold UV luminosity that has been suggested to lower the detection rate of \hi\ absorption at high redshifts. 
The radio luminosities of the AGNs of the sample are also lower than those of a number of AGNs that have yielded detections of \hi\ absorption. Thus, neither the UV nor the radio luminosity of the AGNs of the sample is likely to be the cause of the low detection rate of \hi\ absorption.

Our sample is dominated by extended radio sources. Restricting ourselves to extended radio sources and including five such sources from the literature, and excluding all sources with UV luminosity above $10^{23}$~\wph, we obtain a sample of 28 extended radio sources at $0.7 < z < 1.0$ with $L_{UV} < 10^{23}$~\wph\ with searches for, but no detection of, associated \hi\ absorption. We compare our sample at $0.7 < z < 1.0$ with that of the extended radio sources in \citet{Maccagni17} at $z < 0.25$. Though the detection rates between the two samples appear consistent within $2\sigma$ Gaussian errors, a statistical analysis of the distributions of the \hi\ optical depths of the two samples finds statistically significant ($\sim 3\sigma$) evidence that the strength of \hi\ absorption is lower at high redshifts. The relative low UV and radio luminosities of the AGNs of our sample indicate that this result is unlikely to arise due to a Malmquist bias in the high-$z$ sample. We conclude that our results suggest redshift evolution in the physical conditions of \hii\ in AGN environments at $z < 1$, with high-$z$ AGNs having either high spin temperatures or low \hii\ column densities, resulting in weaker \hi\ absorption.

\begin{acknowledgements}
We thank the referee for useful comments that improved the clarity of the paper. SM thanks Pranav Kukreti for useful discussions. We thank the GMRT staff for making these observations possible. The GMRT is run by the National Centre for Radio Astrophysics of the Tata Institute of Fundamental Research. NK acknowledges support from the Department of Atomic Energy, under project 12-R\&D-TFR-5.02-0700. Funding for the Sloan Digital Sky Survey (SDSS) has been provided by the Alfred P. Sloan Foundation, the Participating Institutions, the National Aeronautics and Space Administration, the National Science Foundation, the U.S. Department of Energy, the Japanese Monbukagakusho, and the Max Planck Society. The SDSS Web site is \url{http://www.sdss.org/}. The SDSS is managed by the Astrophysical Research Consortium (ARC) for the Participating Institutions. The Participating Institutions are The University of Chicago, Fermilab, the Institute for Advanced Study, the Japan Participation Group, The Johns Hopkins University, Los Alamos National Laboratory, the Max-Planck-Institute for Astronomy (MPIA), the Max-Planck-Institute for Astrophysics (MPA), New Mexico State University, University of Pittsburgh, Princeton University, the United States Naval Observatory, and the University of Washington. 

\end{acknowledgements}

%
   \bibliographystyle{aa} 
   \bibliography{ref} 
%


\begin{appendix}
\onecolumn

\section{Tables}
\begin{table}[!h]
\scriptsize
\caption{Observation details.}
\begin{tabular}{ccccccccccccc}
\hline\hline
Source & \textit{z} & $\nu \rm_{obs}$ & $\rm \Delta t$ & BW & $\rm \Delta$v & Beam$_{\rm robust}$ & PA$_{\rm robust}$ & RMS$\rm_{map,robust}$ & Beam$_{\rm natural}$ & PA$_{\rm natural}$ & RMS$_{\rm map,natural}$ & RMS$\rm_{cube}$\\ 
 & & (MHz) & (mins) & (MHz) & (\kmps) & ($^{\prime\prime} \times ^{\prime\prime})$ & ($^{\circ}$) & ($\mu$Jy beam\p{1}) & ($^{\prime\prime} \times ^{\prime\prime}$) & ($^{\circ}$) & ($\mu$Jy beam\p{1}) & (\mjypb) \\ 
 (1) & (2) & (3) & (4) & (5) & (6) & (7) & (8) & (9) & (10) & (11) & (12) & (13)\\
\hline
J000353+121024 & 0.7632   & 805.54  &  265  &  25     & 2.27   &  4.35  $\times$ 3.68  & 46.19   & 75    &  6.12  $\times$  5.43  & -3.74   &  125     &  1.2    \\
J023333-041240 & 0.8786   & 756.09  &  75   & 16.66   & 12.93  &  4.43  $\times$ 3.36  & 49.65   & 163 &  9.71  $\times$  7.86  & 10.74   &  499   &  0.8    \\
J074345+232841 & 0.7764   & 799.56  &  60   & 16.66   & 12.19  &  5.2   $\times$ 3.11  & 68.72   & 233 &  8.98  $\times$  5.4   & 80.6    &  382   &  1.3    \\
J075707+273633 & 0.8183   & 781.14  &  85   & 16.66   & 12.5   &  3.88  $\times$ 3.03  & 44.27   & 118 &  6.78  $\times$  5.9   & 42.53   &  282     &  1.0   \\
J075815+441608 & 0.7808   & 797.58  &  45   & 16.66   & 12.19  &  4.2   $\times$ 2.9   & 40.42   & 135 &  7.88  $\times$  5.02  & 61.32   &  210   &  1.4    \\
J082223+543824 & 0.8588   & 764.13  &  50   & 16.66   & 12.76  &  4.04  $\times$ 2.89  & -2.24   & 100   &  7.53  $\times$  5.25  & 21.83   &  181   &  0.8   \\
J082325+445854 & 0.7956   & 791.04  &  40   & 16.66   & 12.33  &  4.17  $\times$ 3.12  & 29.58   & 148   &  9.35  $\times$  7.86  & 38.15   &  318   &  1.2    \\
J083337+221247 & 0.8098   & 784.80  &  60   & 16.66   & 12.44  &  11.79 $\times$ 4.35  & -87.02  & 221 &  16.38 $\times$  9.86  & -89.94  &   358  &   0.8   \\
J083417+601946 & 0.7167   & 827.38  &  45   & 16.66   & 11.82  &  4.71  $\times$ 3.02  &  7.15   & 131 &  9.58  $\times$  6.85  & 22.53   &  217   &   1.6   \\
J084051+443959 & 0.7693   & 802.76  &  125  &  25     & 2.3    &  5.81  $\times$ 2.78  & 73.92   & 65    &  8.9   $\times$  4.09  & 83.88   &  65      &   1.5   \\
J090835+415046 & 0.7336   & 819.30  &  45   & 16.66   & 11.9   &  4.61  $\times$ 2.78  & 54.56   & 265 &  7.98  $\times$  4.43  & 63.07   &  417   &   2.1   \\
J093000+250005 & 0.7407   & 815.98  &  45   & 16.66   & 11.96  &  6.75  $\times$ 3.1   & 81.05   & 229 &  10.13 $\times$  5.46  & 84.23   &  571     &   1.5   \\
J093150+254034 & 0.8124   & 783.71  &  248  &  25     & 2.3    &  5.62  $\times$ 2.56  & 86.84   & 455   &  7.96  $\times$  3.92  & 85.72   &  560     &   1.3   \\
J101557+010913 & 0.7795   & 798.20  &  55   & 16.66   & 12.19  &  4.54  $\times$ 3.4   & 69.66   & 238 &  8.87  $\times$  7.96  & 76.79   &  795   &   1.2   \\
J110117+331647 & 0.9353   & 733.93  &  65   & 16.63   & 13.27  &  4.33  $\times$ 3.32  & 48.69   & 480 &  9.74  $\times$  5.44  & 56.04   &  856   &   1.5   \\
J110426+492824 & 0.8995   & 747.77  &  35   & 16.66   & 13.02  &  4.34  $\times$ 3.22  & 12.27   & 123 &  9.02  $\times$  6.08  & 44.97   &  208   &   1.3   \\
J110716+053310 & 0.8850    & 753.53  &  10   & 16.66   & 12.93  &  4.96  $\times$ 3.48  & 70.17   & 1684  &  9.47  $\times$  8.72  & 87.83   &  3800    &   3.7   \\
J112723+530058 & 0.9252   & 737.76  &  13   & 25      & 2.48   &  8.4   $\times$ 3.44  & -70.54  & 275   &  9.43  $\times$  6.17  & -71.74  &   535    &   2.5   \\
J113019+101526 & 0.7869   & 794.86  &  80   & 25      & 2.3    &  3.69  $\times$ 3.11  & -7.13   & 200   &  5.91  $\times$  3.71  & 31.99   &  365     &   1.5   \\
J113042+303134 & 0.7367   & 817.86  &  40   & 16.66   & 11.92  &  5.85  $\times$ 3.0     &  75.45  & 214   &  9.27  $\times$  5.42  & 77.91   &  501   &   1.5   \\
J120331+304902 & 0.9557   & 726.26  &  45   & 25      & 2.52   &  4.8   $\times$ 3.31  & -35.46  & 300   &  5.84  $\times$  5.31  & 11.07   &  300     &   1.4   \\
J122451+433519 & 0.9479   & 729.16  &  40   & 25      & 2.5    &  9.29  $\times$ 2.93  &  85.75  & 152 &  11.92 $\times$  4.63  & -89.23  &   173    &   1.2   \\
J131110+343916 & 0.79067  & 793.21  &  270  &  25     & 2.3    &  5.01  $\times$ 3.58  &  82.19  & 40    &  8.73  $\times$  5.3   & 58.17   &  734   &   0.7  \\
J131151+460844 & 0.8284   & 776.84  &  240  &  25     & 2.35   &  6.28  $\times$ 3.17  &  40.92  & 70    &  8.95  $\times$  5.36  & 45.27   &  76      &   0.9  \\
J132754+122308 & 0.9494   & 728.63  &  70   & 25      & 2.51   &  4.63  $\times$ 3.95  & -31.7   & 120   &  6.93  $\times$  5.25  & 39.89   &  170     &   1.0     \\
J160325+143816 & 0.9818  & 716.707 &  40   & 25       & 2.5    &  5.08  $\times$ 4.06  & -39.1   & 190   &  6.7   $\times$  5.21  & 43.43   &  225     &   1.3   \\
J214711+012833 & 0.8811   & 755.08  &  95   & 16.66   & 12.93  &  10.62 $\times$ 3.74  & -32.58  & 527 &  13.14 $\times$  9.4   & 39.73   &  1000    &   1.0  \\
J221029+010843 & 0.7443   & 814.27  &  40   & 16.66   & 11.98  &  4.43  $\times$ 2.88  & 55.05   & 457 &  8.61  $\times$  5.21  & 32.4    &  1000    &   1.9   \\
J225404+005420 & 0.9388   & 732.60  &  30   & 25      & 2.5    &  6.28  $\times$ 3.1   & 48.76   & 210   &  5.97  $\times$  4.84  & 9.97    &  400     &   1.3   \\  
J021755-012150 & 0.9028  & 746.48   &  40  & 16.66    & 12.2 & RFI  \\
J030313-001453 & 0.7000  & 835.53   & 45  & 16.66     & 11.7 & RFI \\
J085448+200630 & 0.7777  & 799.01   & 60  & 25        & 2.29 & RFI \\
\hline
\end{tabular}
\tablefoot{
The columns are: (1) source name; (2) SDSS DR9 redshift; (3) redshifted \hi\ line frequency in MHz; (4) on-source time in minutes; (5) bandwidth of the observations in MHz, (6) spectral resolution in \kmps; (7), (8), and (9) beam size, position angle, and RMS of the continuum image made with ROBUST $-1$ weighting; (10), (11), and (12) beam size, position angle, and RMS of the continuum image made with natural weighting; (13) RMS noise on the cube made with natural weighting with the same beam parameters as the naturally weighted continuum image.
}
\label{radio_obs}
\end{table}

\begin{table*}[!h]
\caption{Derived parameters for the \hi\ non-detections.}
\begin{tabular}{cccccccccccc}
\hline\hline
Source   & $z$  & S$\rm_{\nu}$ (int) & S$\rm_{\nu}$ (peak)  & $\rm \Delta s$    & $\tau_{3\sigma}$  & N$\rm_{HI}$  & L$\rm_{1.4 GHz}$ & L$\rm_{UV}$  & Continuum      \\
& & (mJy) & (mJy beam\p{1}) & (mJy beam\p{1}) & (\%) & (cm\p{2}) & W Hz\p{1} & W Hz\p{1} & morphology\\
(1) & (2) & (3) & (4) & (5) & (6) & (7) & (8) & (9) & (10)\\
\hline
J000353+121024   &   0.7632   &   279.0  &   156.0      &  0.43   &   0.78   &  19.87    &  26.7  &  18.89   &  e  \\
J023333-041240   &   0.8786   &    96.0  &    81.9      &  0.31   &   0.99   &  19.98    &  26.4  &  22.32   &  e  \\
J074345+232841   &   0.7764   &   503.4  &   180.6      &  0.50   &   0.86   &  19.92    &  27.0  &  23.10   &  e  \\
J075707+273633   &   0.8183   &   276.0  &   273.6      &  0.40   &   0.42   &  19.61    &  26.8  &  22.03   &  c  \\
J075815+441608   &   0.7808   &   227.3  &   222.2      &  0.66   &   0.92   &  19.95    &  26.7  &  18.14   &  c  \\
J082223+543824   &   0.8588   &   346.9  &   267.7      &  0.35   &   0.29   &  19.45    &  26.9  &  21.61   &  e  \\
J082325+445854   &   0.7956   &   574.3  &   82.0$^*$   &  0.56   &   2.19   &  20.32    &  27.1  &  22.60   &  e  \\
J083337+221247   &   0.8098   &   299.4  &   211.8      &  0.29   &   0.42   &  19.60    &  26.8  &  22.14   &  e  \\
J083417+601946   &   0.7167   &   189.6  &   160.2      &  0.53   &   0.84   &  19.91    &  26.5  &  22.12   &  e  \\
J084051+443959   &   0.7693   &   136.2  &   135.6      &  0.52   &   1.06   &  20.01    &  26.4  &  22.24   &  c  \\
J090835+415046   &   0.7336   &   294.0  &   218.7      &  0.62   &   0.85   &  19.91    &  26.6  &  22.59   &  e  \\
J093000+250005   &   0.7407   &   593.7  &   474.3      &  0.61   &   0.32   &  19.49    &  27.0  &  22.49   &  e  \\
J093150+254034   &   0.8124   &   203.0  &   202.0      &  0.51   &   0.67   &  19.81    &  26.7  &  21.45   &  c  \\
J101557+010913   &   0.7795   &    97.8  &    98.7      &  0.55   &   1.52   &  20.16    &  26.3  &  23.73   &  c  \\
J110117+331647   &   0.9353   &   283.6  &   251.0      &  0.62   &   0.73   &  19.85    &  26.9  &  22.80   &  e  \\
J110426+492824   &   0.8995   &   355.2  &   349.6      &  0.56   &   0.43   &  19.62    &  27.0  &  22.05   &  c  \\
J110716+053310   &   0.885    &   925.4  &   370.0      &  1.48   &   1.16   &  20.05    &  27.4  &  22.80   &  e  \\
J112723+530058   &   0.9252   &   358.0  &   304.0      &  1.33   &   0.90   &  19.94    &  27.0  &  21.96   &  e  \\
J113019+101526   &   0.7869   &   578.3  &   367.0      &  0.66   &   0.54   &  19.72    &  27.1  &  21.92   &  e  \\
J113042+303134   &   0.7367   &   586.4  &   381.9      &  0.66   &   0.55   &  19.72    &  27.0  &  22.83   &  e  \\
J120331+304902   &   0.9557   &   190.0  &   173.0      &  0.75   &   1.18   &  20.05    &  26.8  &  22.39   &  e  \\
J122451+433519   &   0.9479   &   460.9  &   394.4      &  0.63   &   0.39   &  19.58    &  27.2  &  22.06   &  e  \\
J131110+343916   &   0.7906   &   153.9  &   127.4      &  0.31   &   0.74   &  19.85    &  26.5  &  20.02   &  e  \\
J131151+460844   &   0.8284   &   215.9  &   182.3      &  0.37   &   0.64   &  19.79    &  26.7  &  20.22   &  e  \\
J132754+122308   &   0.9494   &   604.0  &   148.0$^*$  &  0.44   &   1.00   &  19.98    &  27.3  &  22.12   &  e  \\
J160325+143816   &   0.9818   &   379.1  &   372.7      &  0.62   &   0.57   &  19.74    &  27.1  &  21.83   &  c  \\
J214711+012833   &   0.8811   &   179.9  &   143.5      &  0.40   &   0.77   &  19.87    &  26.7  &  20.58   &  e  \\
J221029+010843   &   0.7443   &   324.3  &   319.8      &  0.74   &   0.65   &  19.80    &  26.8  &  20.80   &  c  \\
J225404+005420   &   0.9388   &   647.4  &   581.5      &  0.59   &   0.32   &  19.49    &  27.3  &  21.92   &  e  \\

\hline
\end{tabular}
\tablefoot{
The columns are: (1) source name; (2) SDSS DR9 redshift; (3) integrated radio continuum flux in mJy; (4) peak flux density in \mjypb; (5) RMS noise on the spectrum after smoothing it to around 50 \kmps, in \mjypb; (6) 3$\sigma$ optical depth limit assuming c$_f =0.5$; (7) 3$\sigma$ limit on the \hii\ column density in log scale estimated assuming a \tspin $\rm= 100$ K; (8) rest-frame 1.4 GHz radio luminosity in log scale; (9) rest-frame UV luminosity in log scale; (10) continuum morphology as seen in our observations  -- e: extended, c: compact. Sources with a ratio S$_\nu$(peak)/S$_\nu$(int) $> 0.9$ were considered compact and the rest extended sources. The UV luminosity was estimated using the SDSS u- and g-band magnitudes.\\ $^*$In this case the optical image showed that the radio core does not overlap with the peak radio emission. So we extracted the flux density from the beam element corresponding to the location of the radio core.\\ 
}
\label{radio_table_nondet}
\end{table*}

\clearpage
\newpage

\section{\hii\ spectra}
\label{sec:HIundetected}
\begin{figure*}[!h]
\includegraphics[width=6cm, height=4cm]{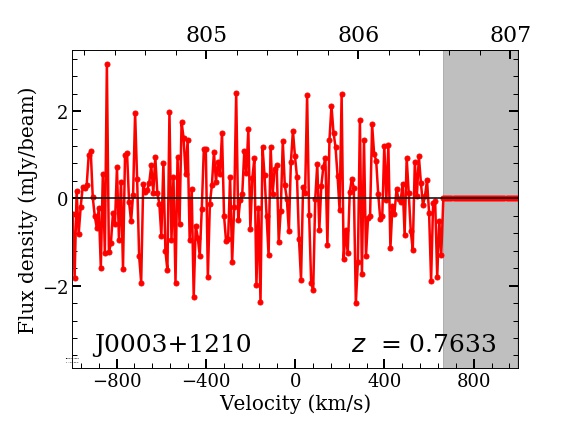}
\includegraphics[width=6cm, height=4cm]{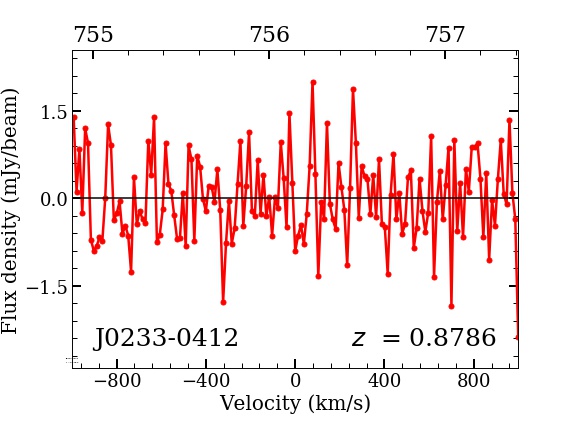}
\includegraphics[width=6cm, height=4cm]{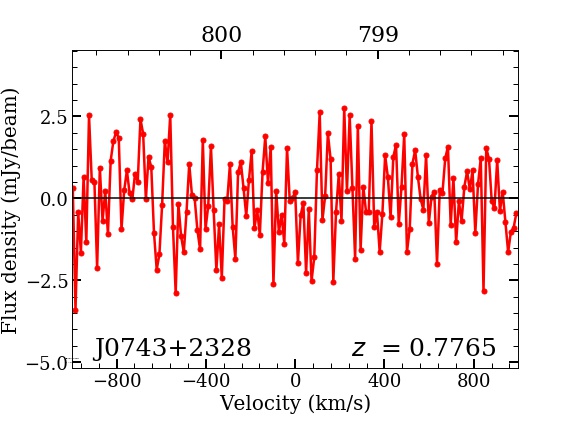}
\includegraphics[width=6cm, height=4cm]{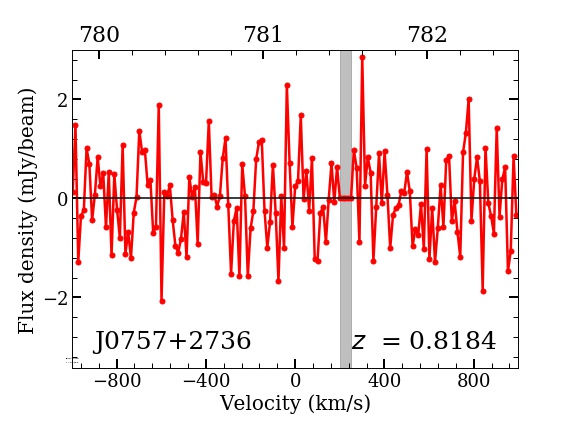}
\includegraphics[width=6cm, height=4cm]{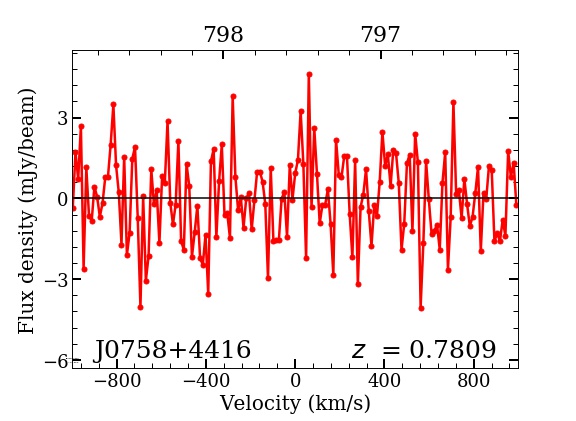}
\includegraphics[width=6cm, height=4cm]{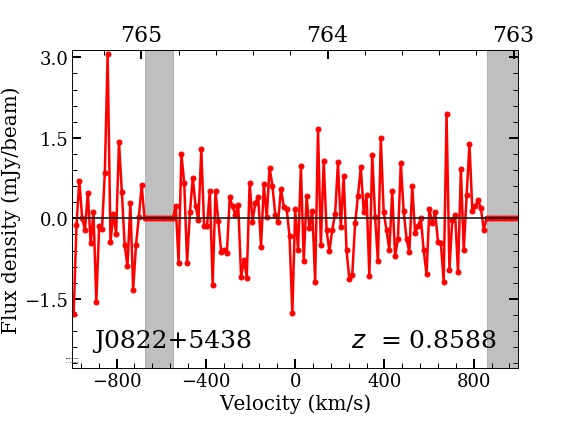}
\includegraphics[width=6cm, height=4cm]{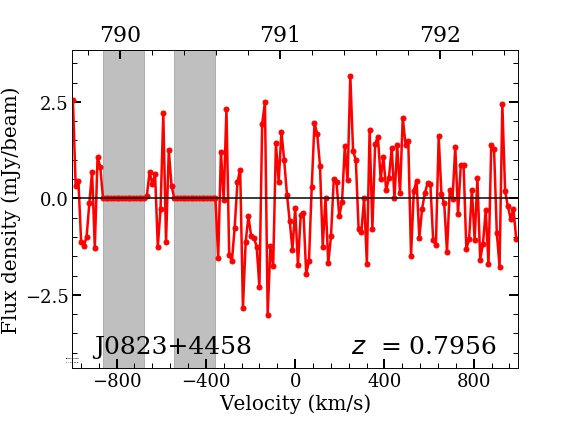}
\includegraphics[width=6cm, height=4cm]{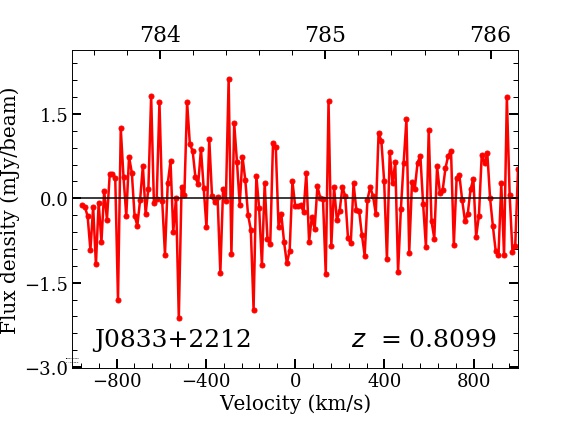}
\includegraphics[width=6cm, height=4cm]{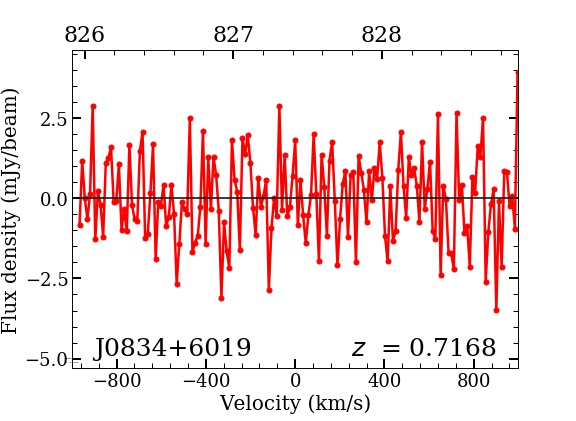}
\includegraphics[width=6cm, height=4cm]{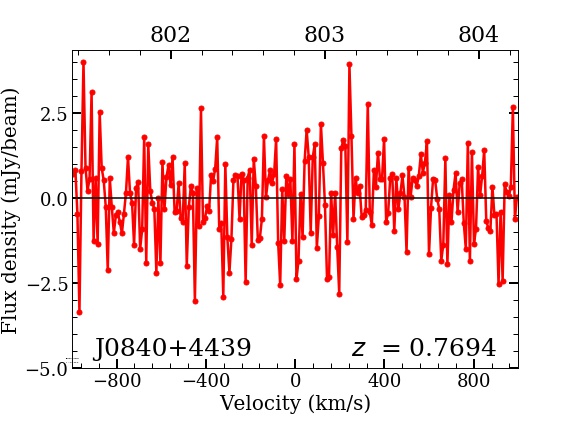}
\includegraphics[width=6cm, height=4cm]{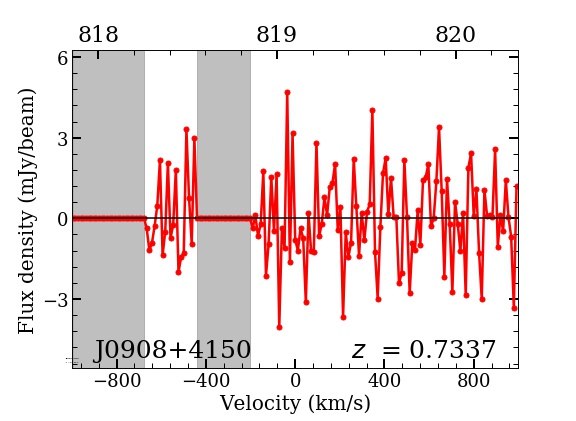}
\includegraphics[width=6cm, height=4cm]{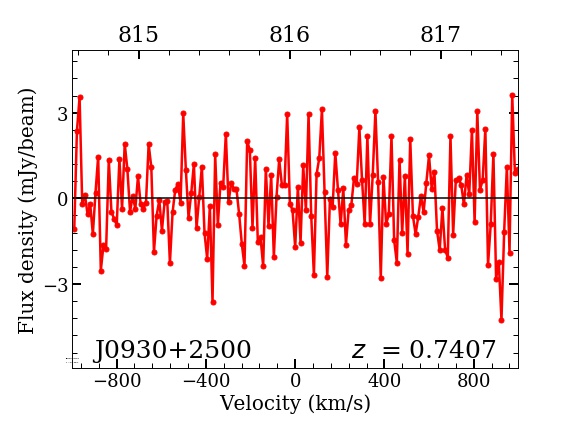}
\includegraphics[width=6cm, height=4cm]{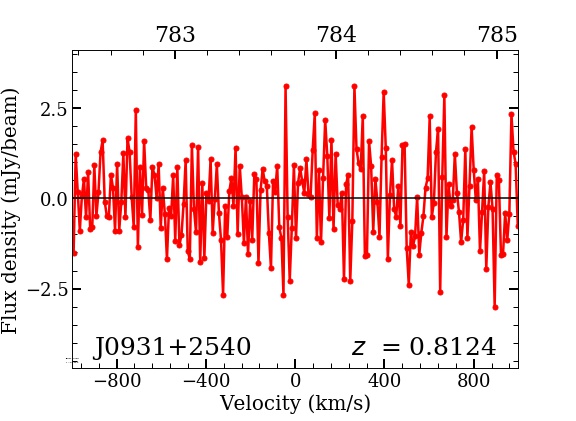}
\hspace{0.1em}
\includegraphics[width=6cm, height=4cm]{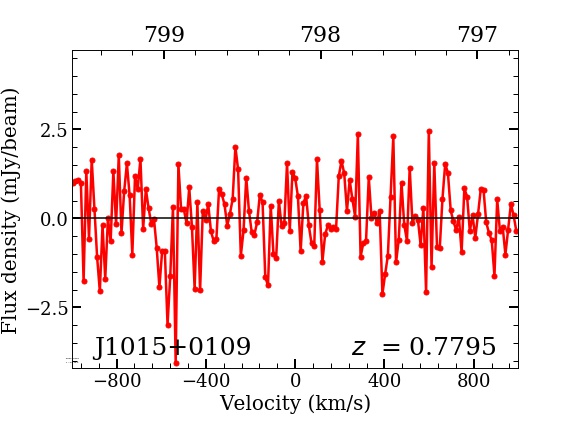}
\hspace{0.1em}
\includegraphics[width=6cm, height=4cm]{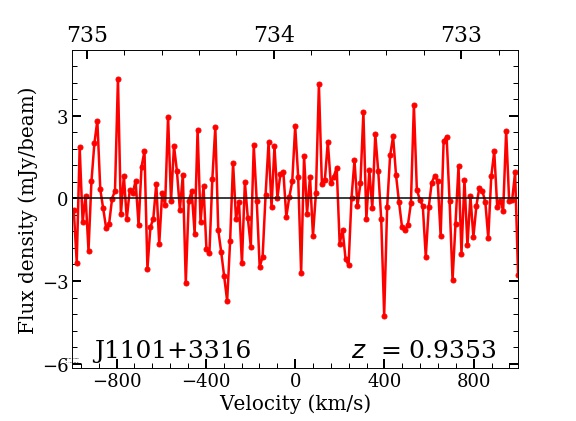}
   \caption{Spectra of the non-detections. The heliocentric frequency, in MHz, is marked on top of each panel. The velocity ranges affected by RFI are shown as grey shaded regions.}
    \label{fig:nondet_spectra}
\end{figure*}

\begin{figure*}[!h]\ContinuedFloat
\includegraphics[width=6cm, height=4cm]{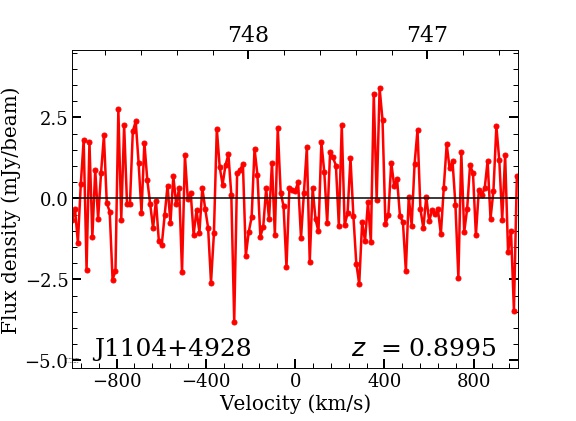}
\includegraphics[width=6cm, height=4cm]{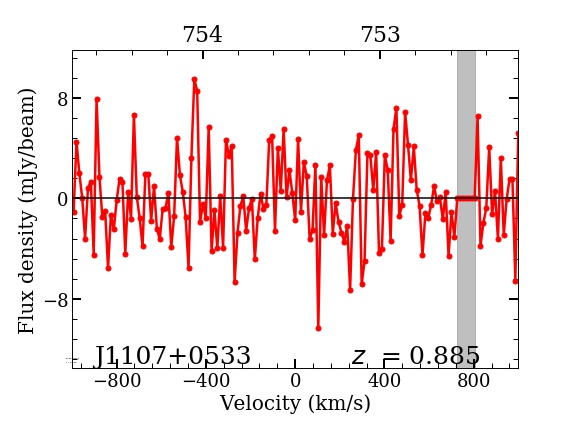}
\includegraphics[width=6cm, height=4cm]{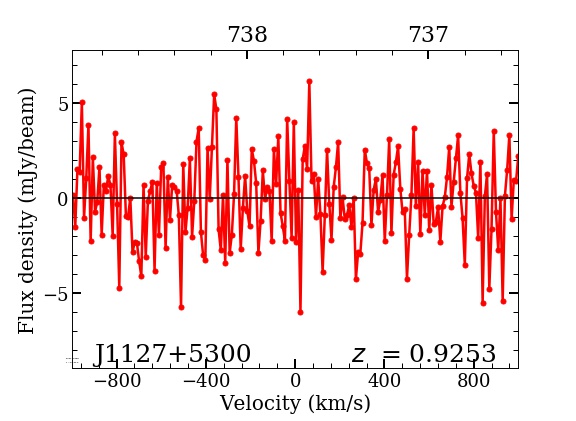}
\includegraphics[width=6cm, height=4cm]{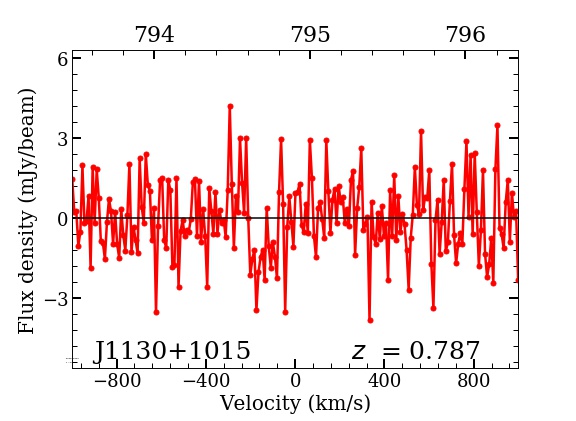}
\includegraphics[width=6cm, height=4cm]{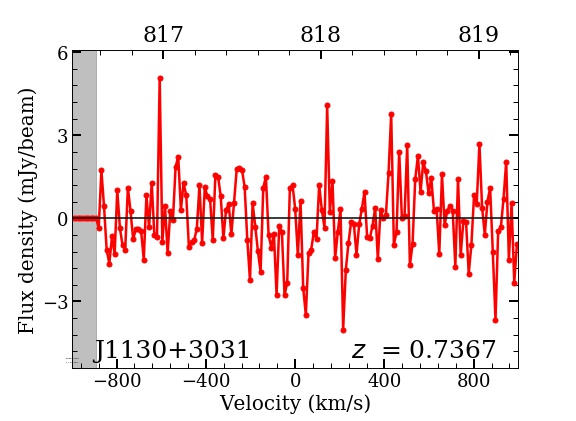}
\includegraphics[width=6cm, height=4cm]{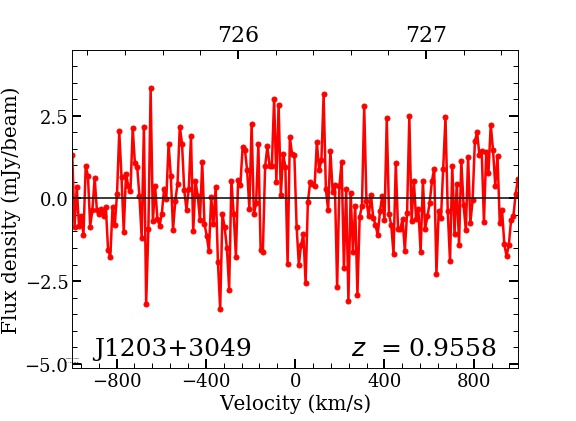}
\includegraphics[width=6cm, height=4cm]{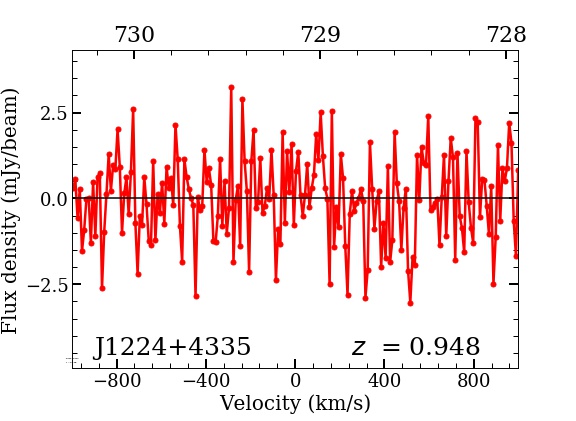}
\includegraphics[width=6cm, height=4cm]{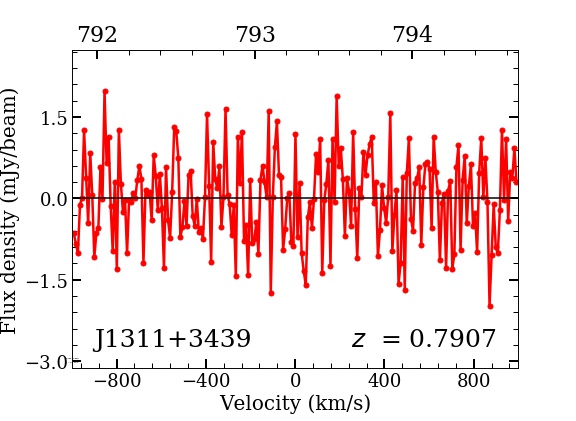}
\includegraphics[width=6cm, height=4cm]{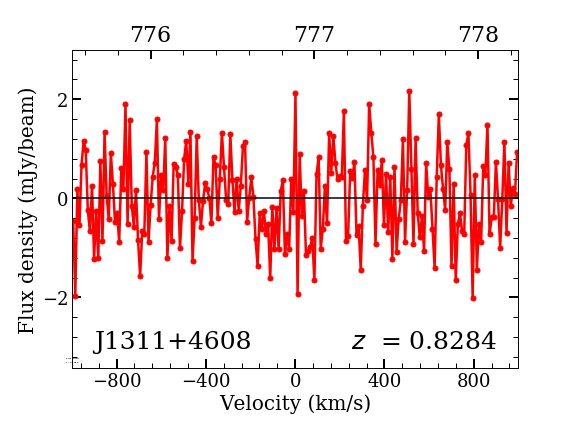}
\includegraphics[width=6cm, height=4cm]{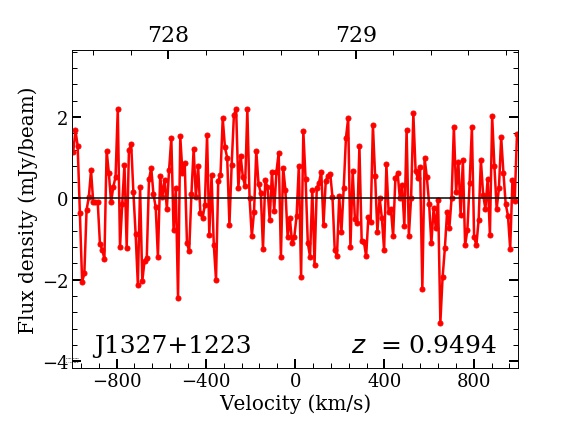}
\includegraphics[width=6cm, height=4cm]{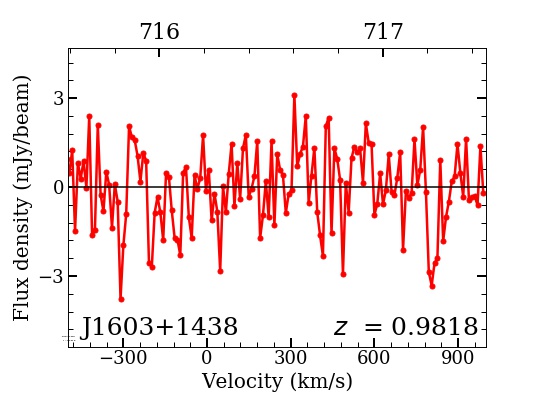}
\includegraphics[width=6cm, height=4cm]{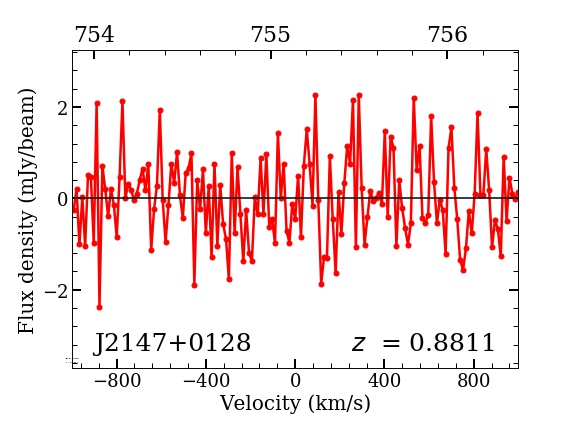}
\hspace{0.1em}
\includegraphics[width=6cm, height=4cm]{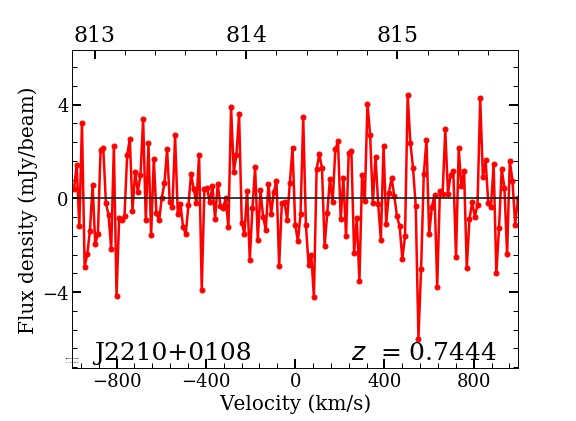}
\hspace{0.1em}
\includegraphics[width=6cm, height=4cm]{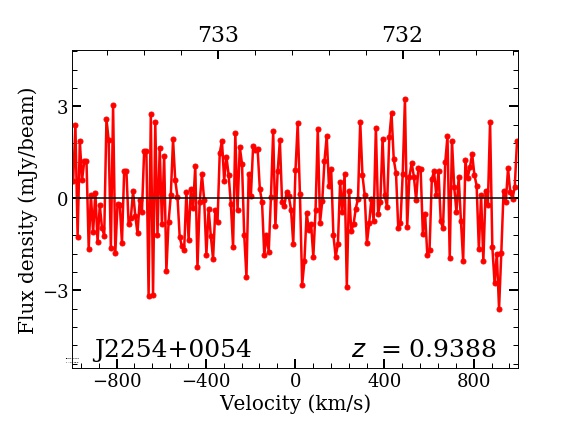}

   \caption{continued.}
    \label{fig:nondet_spectra}
\end{figure*}

\clearpage
\newpage
\section{Continuum images of the radio sources}
\label{sec:ContImages}

\begin{figure*}[!h]

\includegraphics[width=6cm, height=4.3cm]{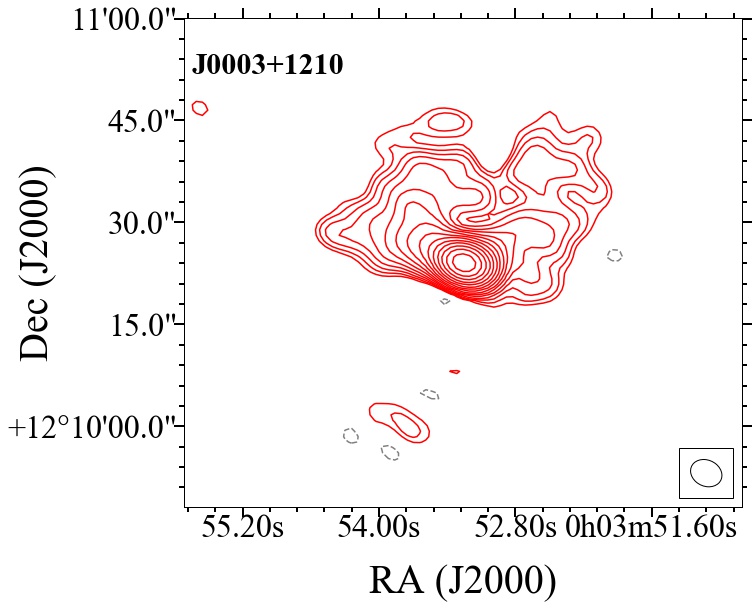}
\includegraphics[width=6cm, height=4.3cm]{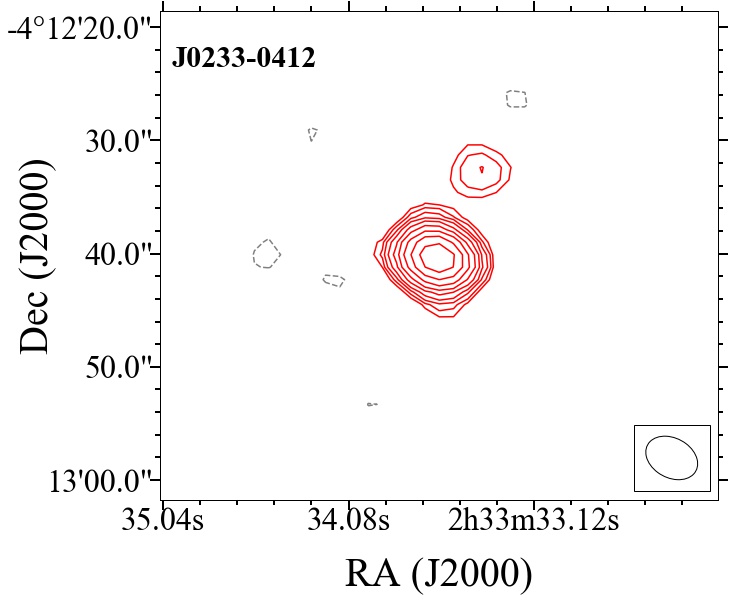}
\includegraphics[width=6cm, height=4.3cm]{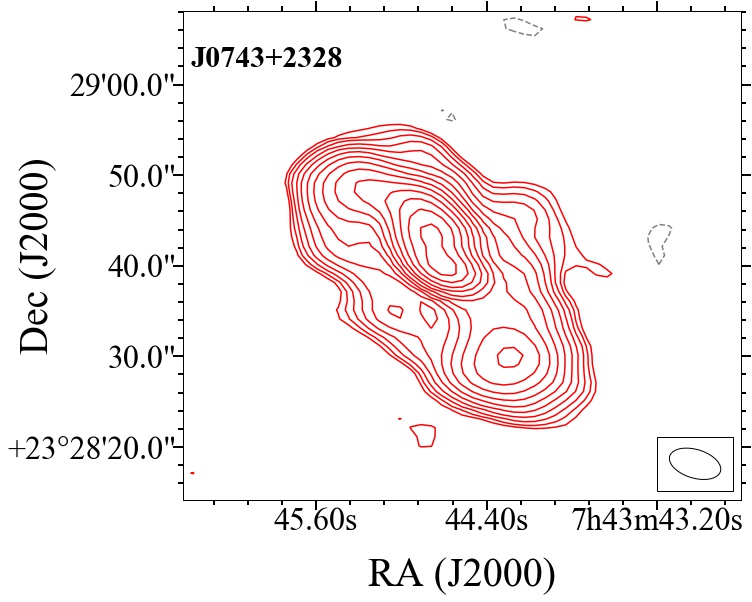}
\includegraphics[width=6cm, height=4.3cm]{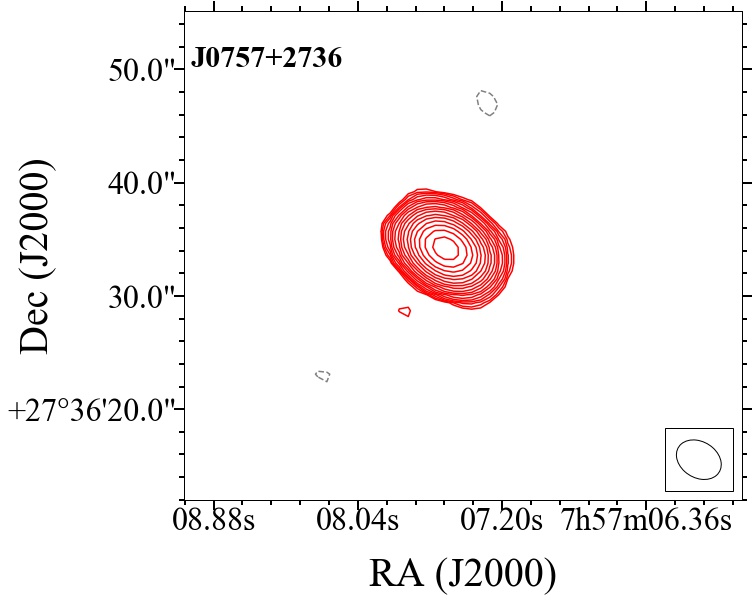}
\includegraphics[width=6cm, height=4.3cm]{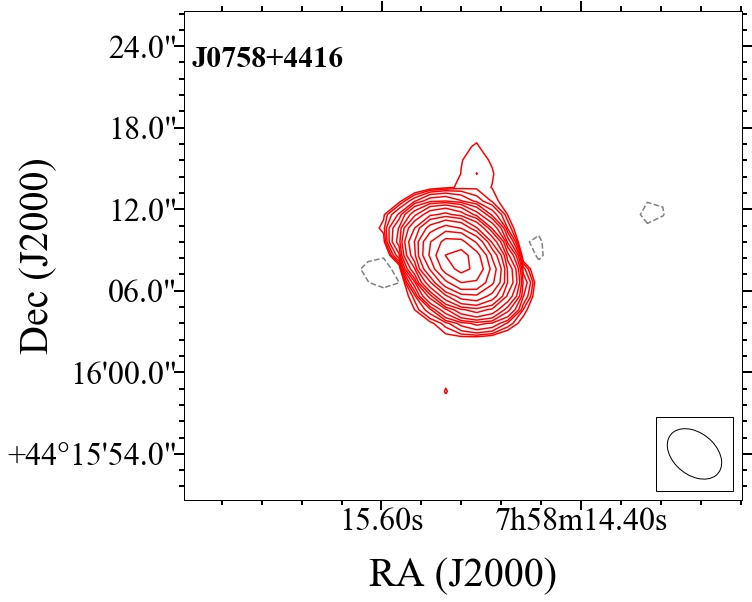}
\includegraphics[width=6cm, height=4.3cm]{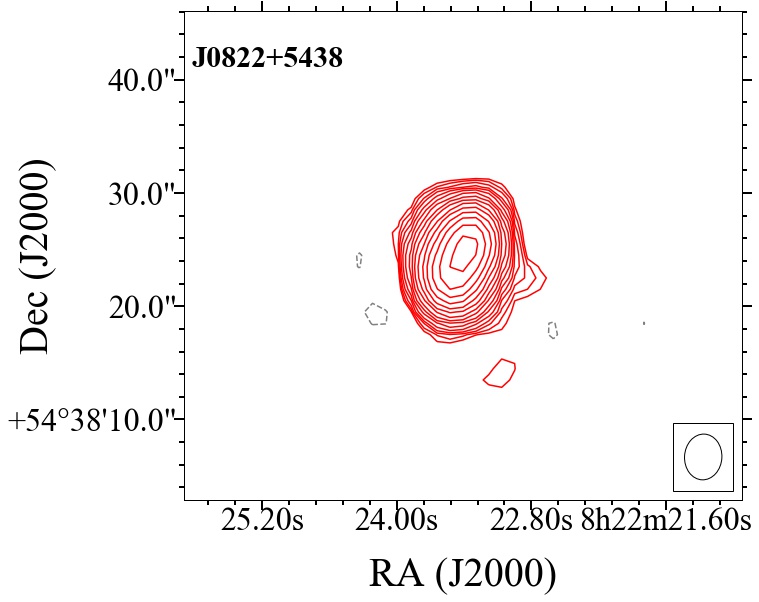}
\includegraphics[width=6cm, height=4.3cm]{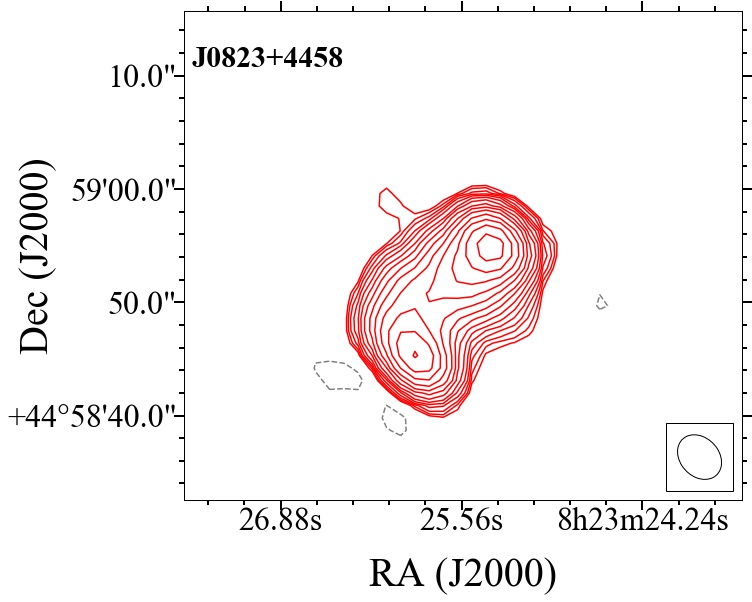}
\includegraphics[width=6cm, height=4.3cm]{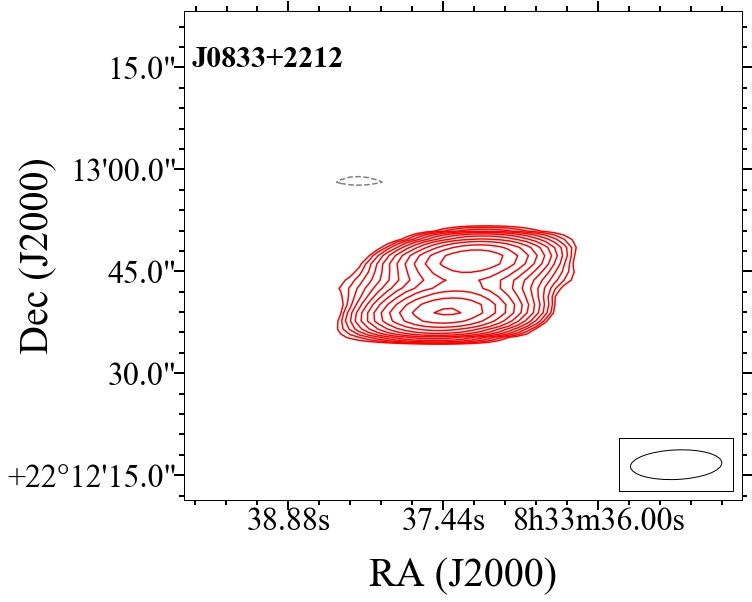}
\includegraphics[width=6cm, height=4.3cm]{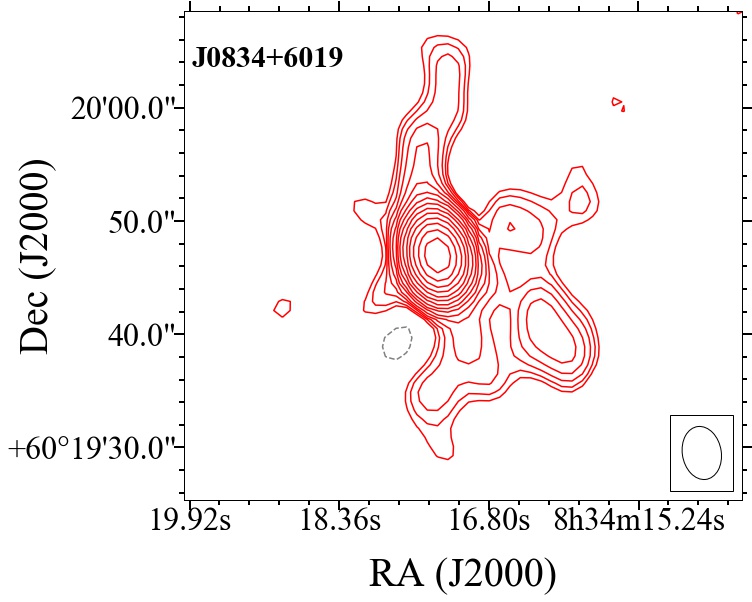}
\includegraphics[width=6cm, height=4.3cm]{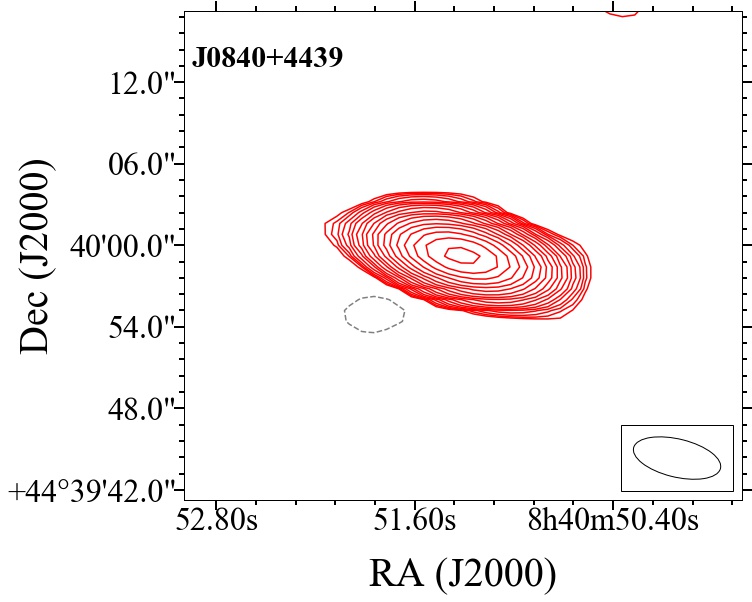}
\hspace{0.1em}
\includegraphics[width=6cm, height=4.3cm]{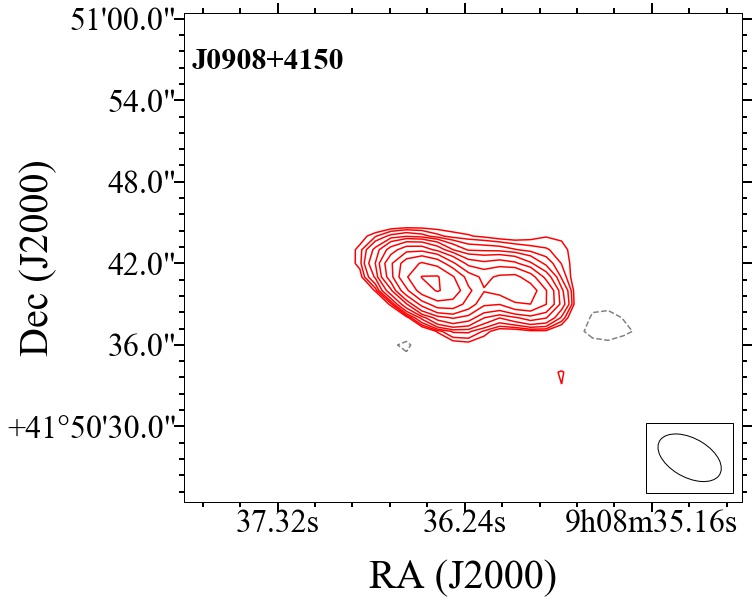}
\hspace{0.1em}
\includegraphics[width=6cm, height=4.3cm]{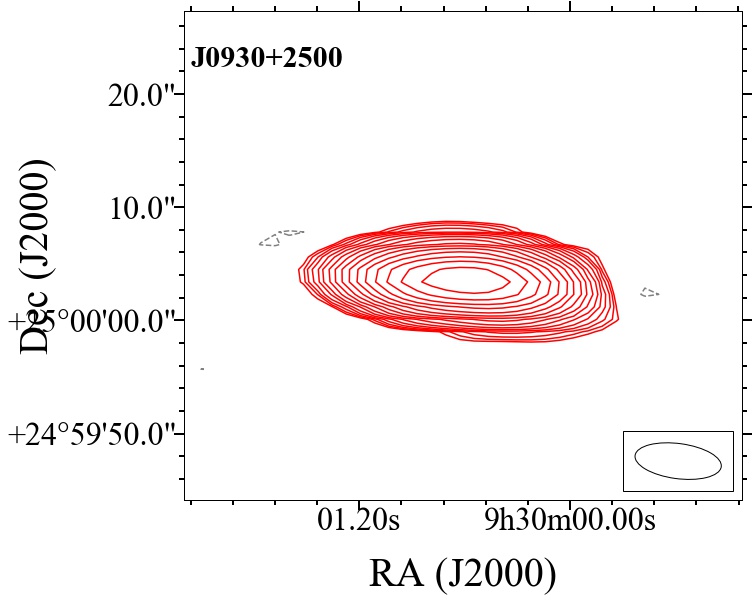}

\caption{Radio continuum images of the sources in our sample. The contours typically start at 4$\sigma$ and increase in steps of $\sqrt{2}$. The RMS noise and the beam sizes are listed in Table \ref{radio_table_nondet}. The 4$\sigma$ negative contours are shown in grey.}
\label{fig:continuum_maps}
\end{figure*}

\begin{figure*}
\ContinuedFloat
\includegraphics[width=6cm, height=4.3cm]{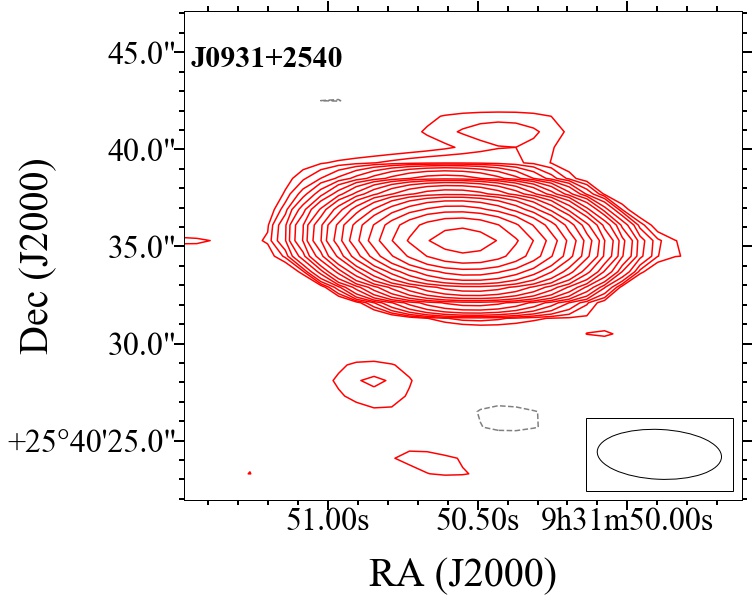}
\includegraphics[width=6cm, height=4.3cm]{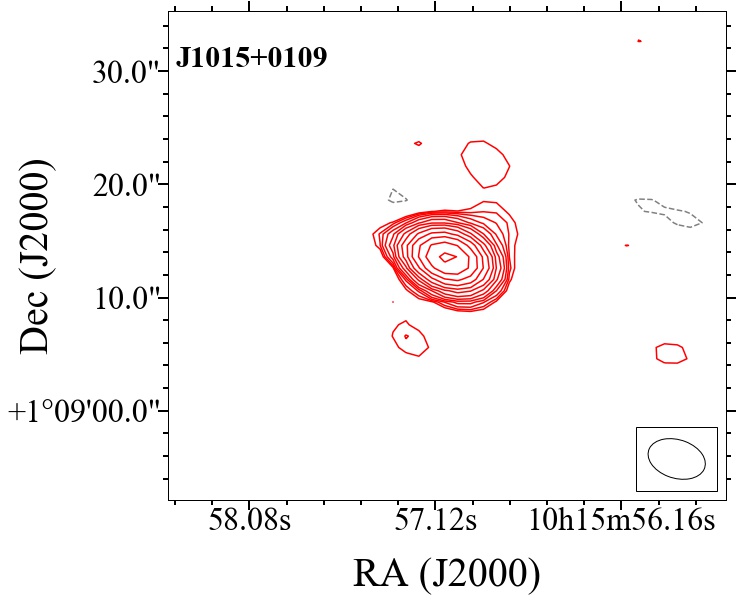}
\includegraphics[width=6cm, height=4.3cm]{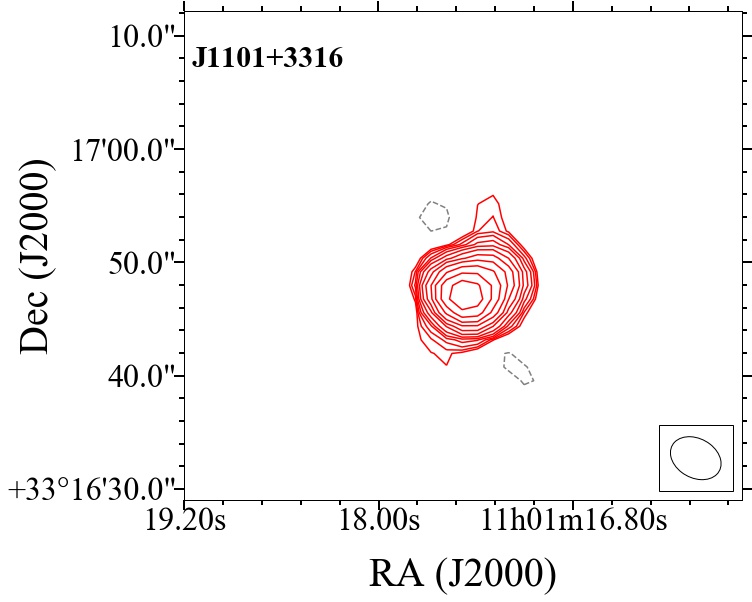}
\includegraphics[width=6cm, height=4.3cm]{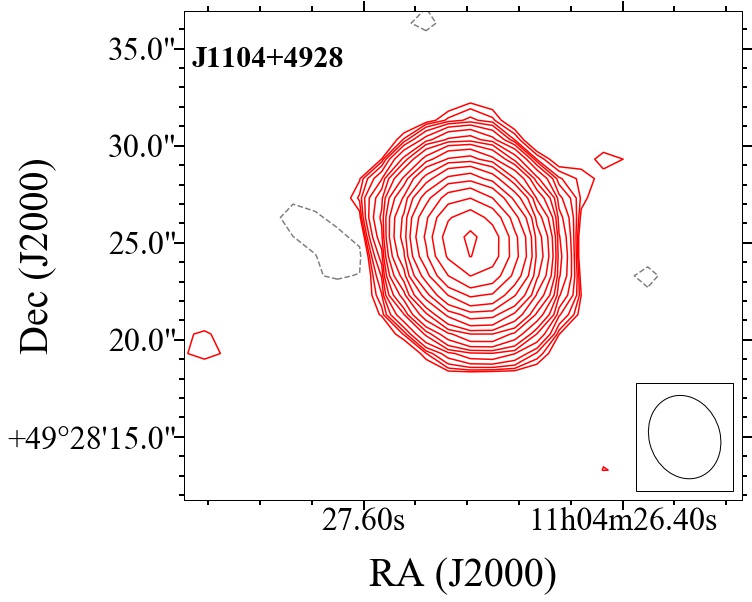}
\includegraphics[width=6cm, height=4.3cm]{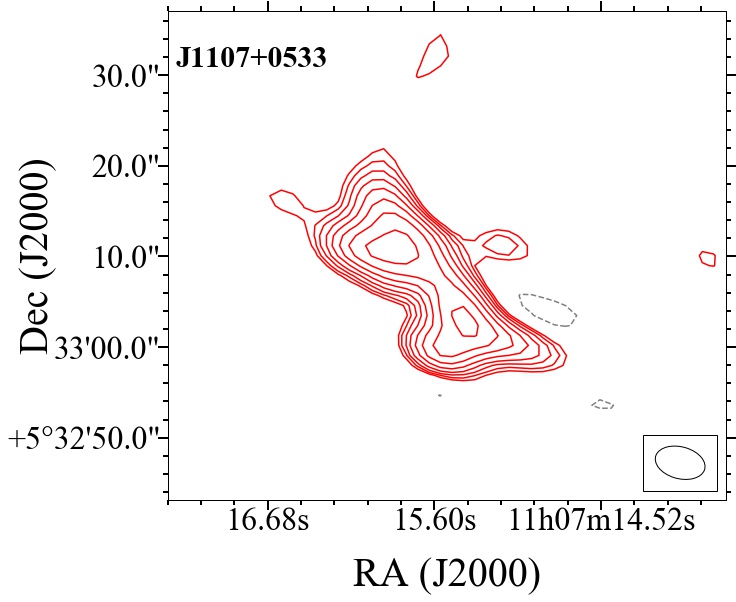}
\includegraphics[width=6cm, height=4.3cm]{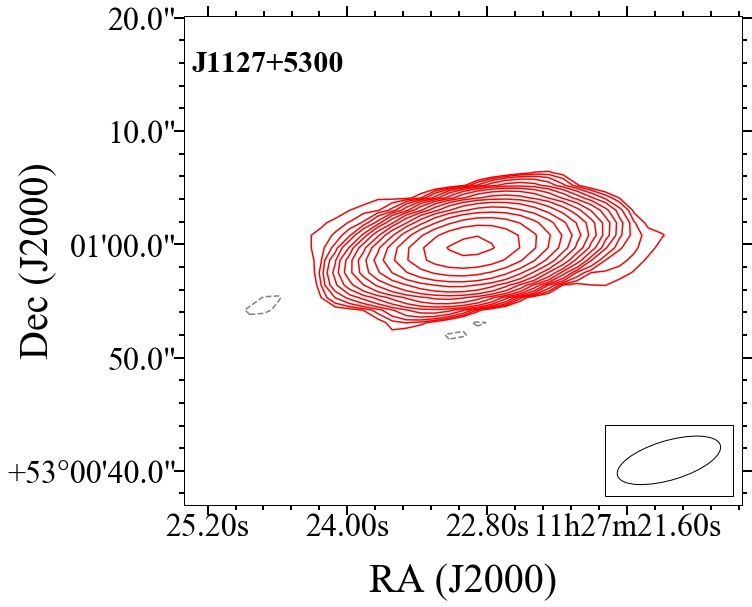}
\includegraphics[width=6cm, height=4.3cm]{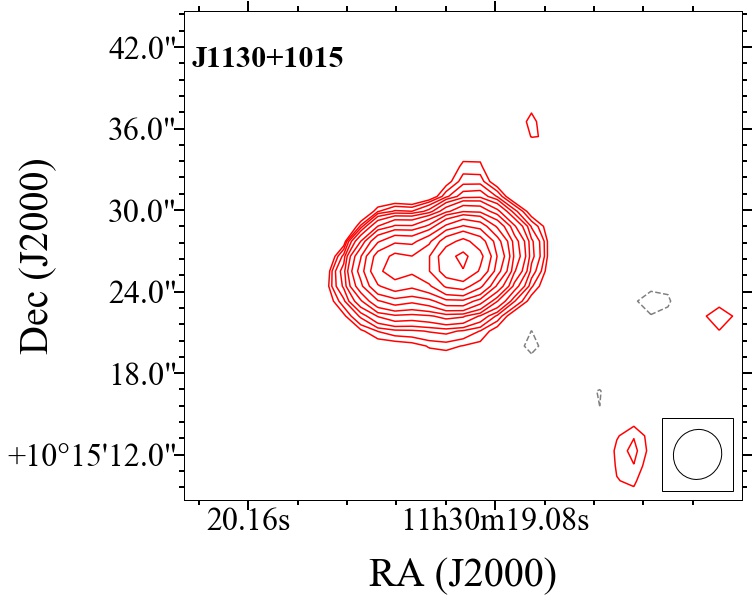}
\includegraphics[width=6cm, height=4.3cm]{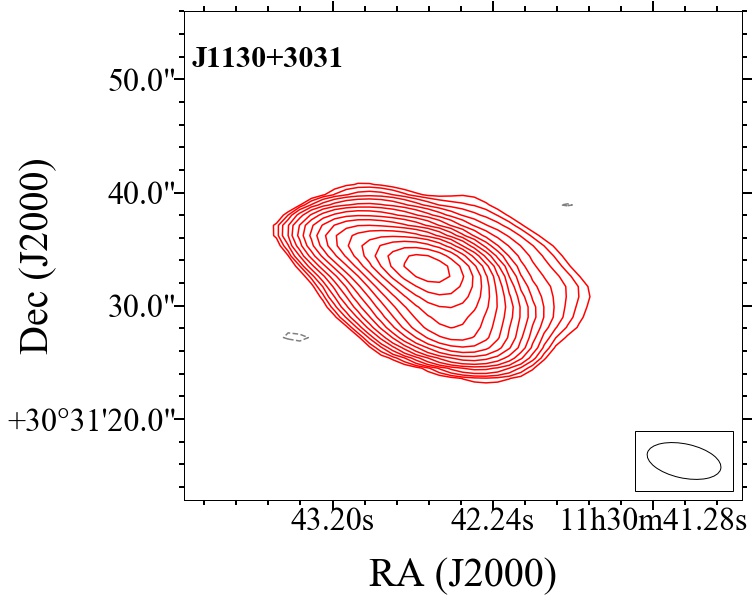}
\includegraphics[width=6cm, height=4.3cm]{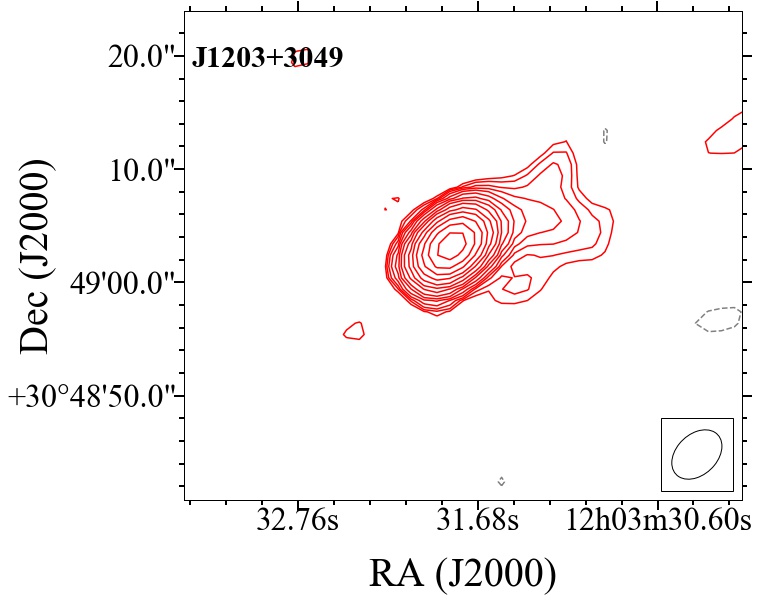}
\includegraphics[width=6cm, height=4.3cm]{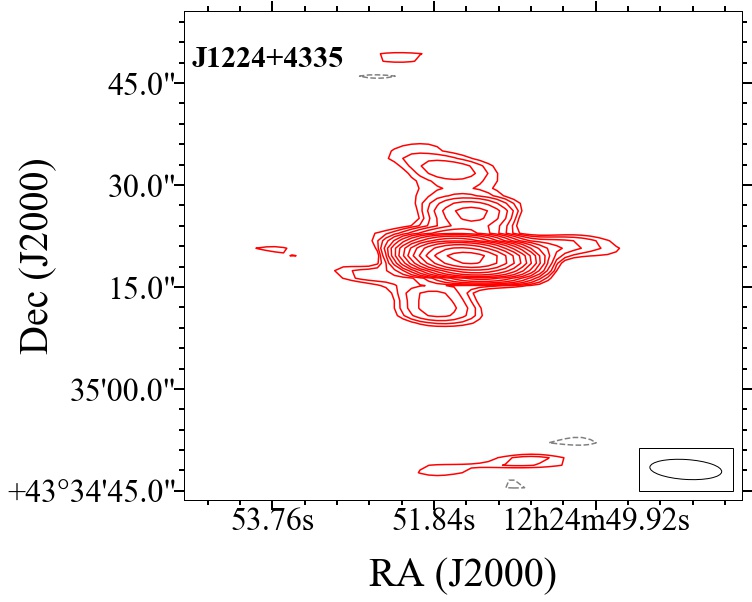}
\includegraphics[width=6cm, height=4.3cm]{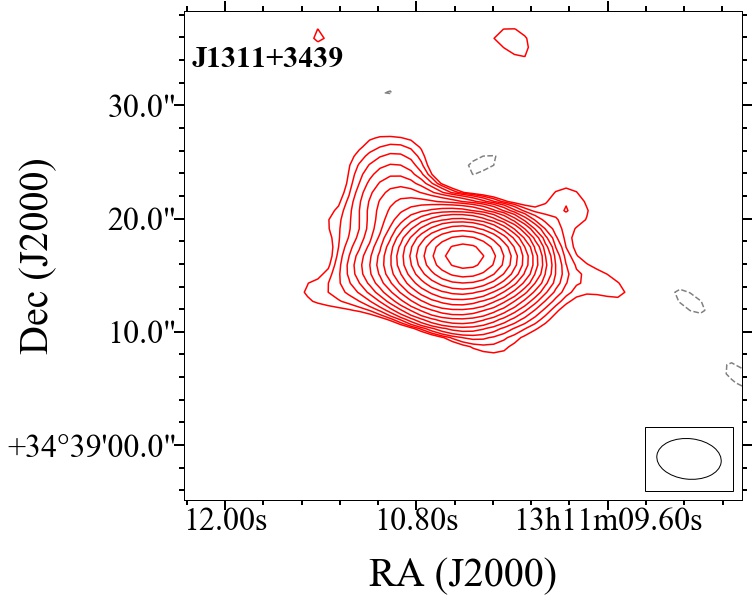}
\includegraphics[width=6cm, height=4.3cm]{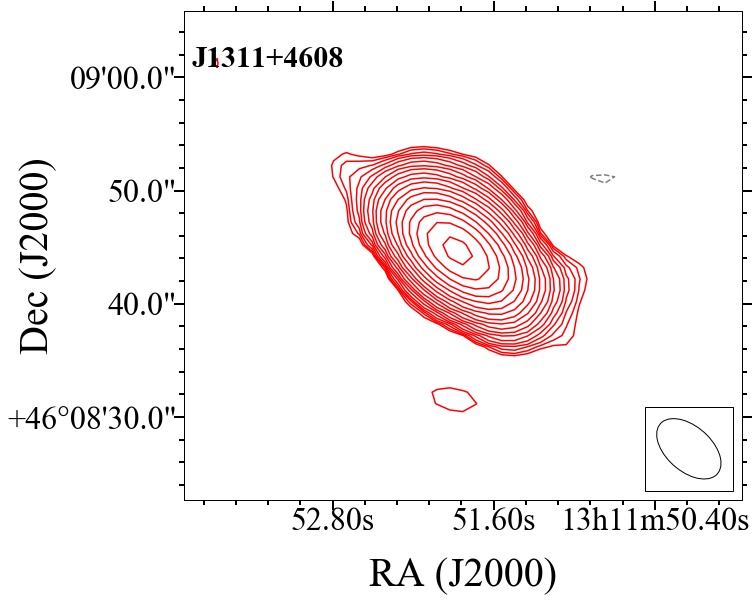}
\includegraphics[width=6cm, height=4.3cm]{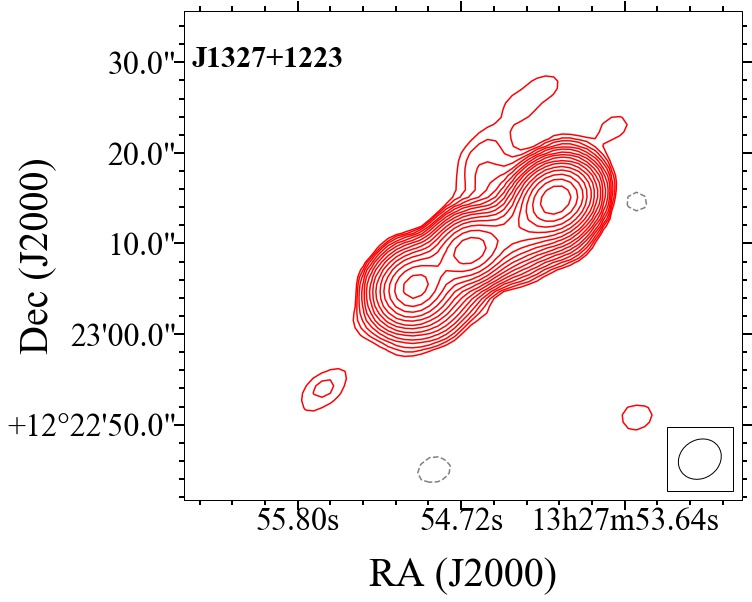}
\hspace{0.1em}
\includegraphics[width=6cm, height=4.3cm]{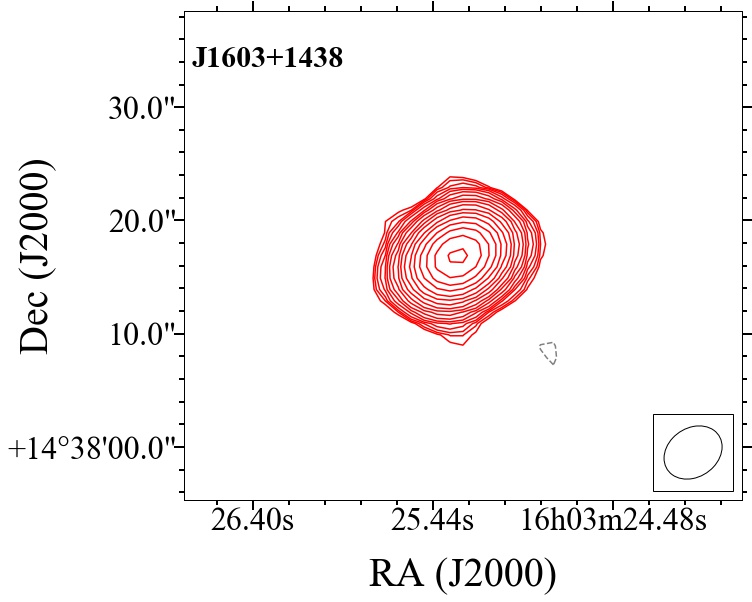}
\hspace{0.1em}
\includegraphics[width=6cm, height=4.3cm]{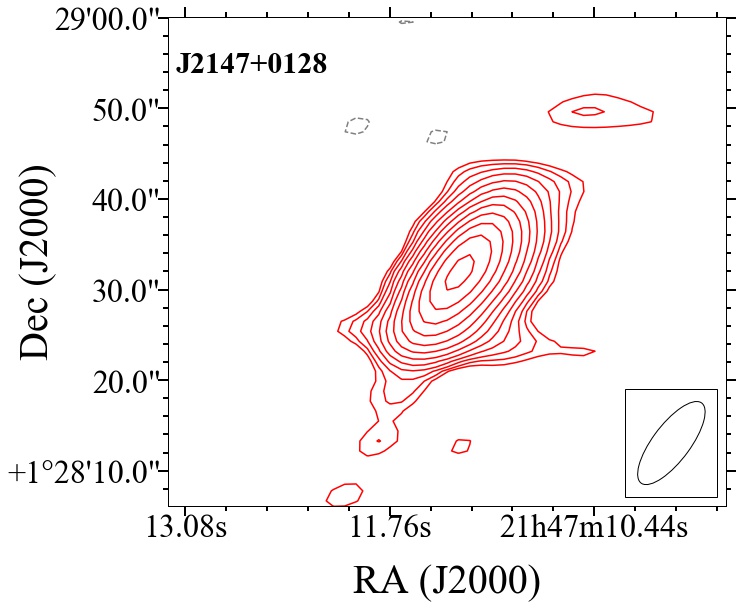}
   \caption{continued.}
    \label{fig:continuum_maps}
\end{figure*}

\begin{figure*}
\ContinuedFloat
\includegraphics[width=6cm, height=4.3cm]{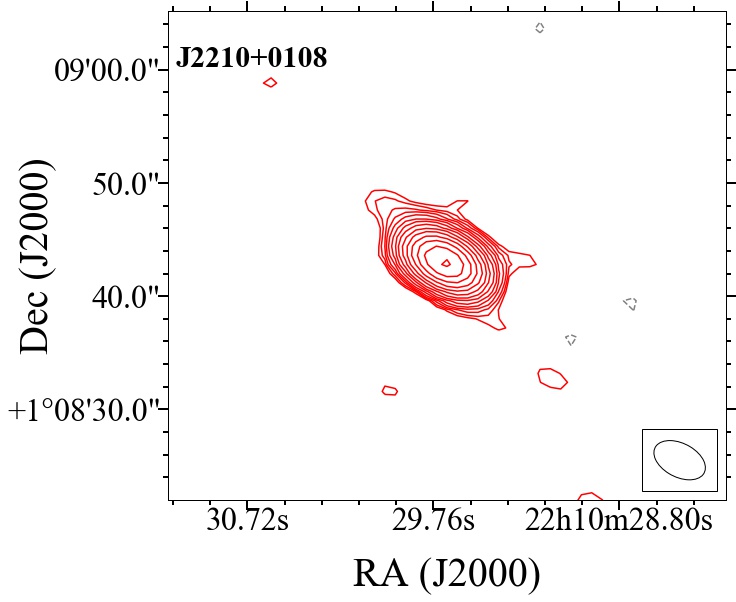}
\includegraphics[width=6cm, height=4.3cm]{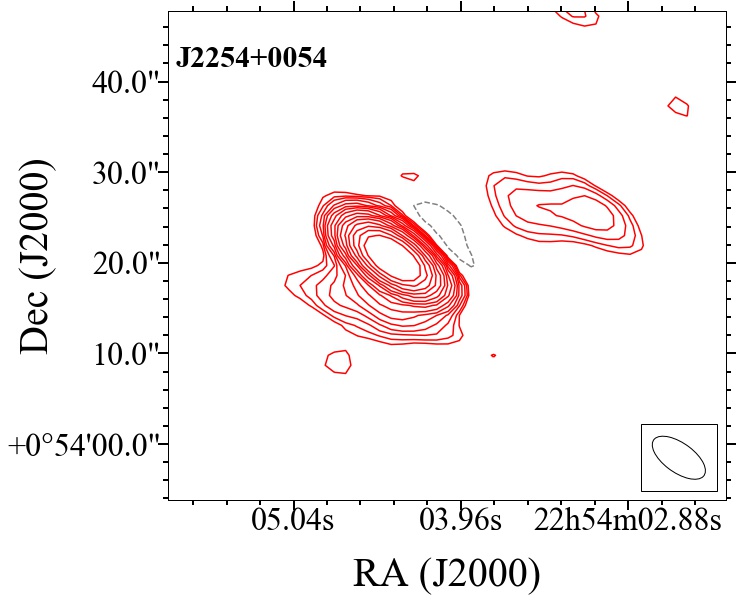}
   \caption{continued.}
    \label{fig:continuum_maps}
\end{figure*}

\end{appendix}


\end{document}